\documentclass[sigconf]{acmart}
\newcommand{\tuner}{\texttt{PromiseTune}}

\usepackage{hyperref}
\usepackage[ruled, linesnumbered, vlined, commentsnumbered]{algorithm2e}
\usepackage{amsmath,blkarray,booktabs}
\usepackage{xcolor,colortbl}
\usepackage{multirow}
\usepackage{subcaption} 
\usepackage{tcolorbox}
\usepackage{csquotes}
\usepackage{tikz,pgfplots}
\usepackage{standalone}
\usepackage{pgfplots}
\usepackage{graphicx}
\usepackage{balance}
\usepackage{pifont}
\usepackage{algorithmic} 
\usepackage{tikz}
\usepackage[ruled,vlined,linesnumbered]{algorithm2e}
\usepackage{color}
\usepackage{pifont}
\newtcolorbox{quotebox}{colback=blue!10,boxrule=0.4pt,colframe=black,fonttitle=\bfseries,top=2pt,bottom=2pt}

{\authoralign{#1}
\ifblank{#2}
   {\def\shadequoteauthor{}\def\yshift{-2ex}\def\quotefill{\hfill}}
   {\def\shadequoteauthor{\par\authorfill\shadedauthorformat{#2}}\def\yshift{1ex}}
\begin{snugshade}\begin{quote}\openquote}
{\shadequoteauthor\quotefill\closequote{\yshift}\end{quote}\end{snugshade}}

\newcommand{\vect}[1]{{\boldsymbol{#1}}}

\DeclareMathAlphabet\mathbfcal{OMS}{cmsy}{b}{n}

  \newcommand{\quart}[4]{\begin{adjustbox}{max width=.1\textwidth}\begin{picture}(10,5)
    {\color{ultramarine}\put(5,2.5){\circle*{5}}\color{black}\put(5,2.5){\circle{5}}}\end{picture}\end{adjustbox}}
\AtBeginDocument{%
  }

\definecolor{ultramarine}{RGB}{0,128,1}

\copyrightyear{2026}
\acmYear{2026}
\setcopyright{rightsretained}
\acmConference[ICSE '26]{2026 IEEE/ACM 48th International Conference on Software Engineering}{April 12--18, 2026}{Rio de Janeiro, Brazil}
\acmBooktitle{2026 IEEE/ACM 48th International Conference on Software Engineering (ICSE '26), April 12--18, 2026, Rio de Janeiro, Brazil}
\acmPrice{}
\acmDOI{10.1145/3744916.3764552}
\acmISBN{979-8-4007-2025-3/26/04}

\begin{document}


\title{PromiseTune: Unveiling Causally Promising and Explainable Configuration Tuning}

\author{Pengzhou Chen}
\email{cc15523016531@gmail.com}
\authornote{Pengzhou Chen is also supervised in the IDEAS Lab.}
\affiliation{%
  \institution{School of Computer Science and Engineering\\University of Electronic Science and Technology of China}
  \city{Chengdu}
  \country{China}
}

\author{Tao Chen}
\email{t.chen@bham.ac.uk}
\authornote{Tao Chen is the corresponding author.}
\affiliation{%
  \institution{IDEAS Lab, School of Computer Science\\University of Birmingham}
  \city{Birmingham}
  \country{UK}
}


\begin{abstract}
    The high configurability of modern software systems has made configuration tuning a crucial step for assuring system performance, e.g., latency or throughput. However, given the expensive measurements, large configuration space, and rugged configuration landscape, existing tuners suffer ineffectiveness due to the difficult balance of budget utilization between exploring uncertain regions (for escaping from local optima) and exploiting guidance of known good configurations (for fast convergence). The root cause is that we lack knowledge of where the \textbf{\textit{promising regions}} lay, which also causes challenges in the explainability of the results.

    In this paper, we propose \tuner~that tunes the configuration guided by causally purified rules. \tuner~is unique in the sense that we learn rules, which reflect certain regions in the configuration landscape, and purify them with causal inference. The remaining rules serve as approximated reflections of the promising regions, bounding the tuning to emphasize these places in the landscape. This, as we demonstrate, can effectively mitigate the impact of the exploration and exploitation trade-off. Those purified regions can then be paired with the measured configurations to provide spatial explainability at the landscape level. Compared with 11 state-of-the-art tuners on 12 systems and varying budgets, we show that \tuner~performs significantly better than the others with $42\%$ superior rank to the overall second best while providing richer information to explain the hidden system characteristics.
    
\end{abstract}




\maketitle

\section{Introduction}
\label{sec:introduction}

Software systems are becoming increasingly configurable, providing great flexibility to software users. However, this also incurs the difficulty of how to tune the configuration since it can profoundly impact system performance, e.g., latency and throughput~\cite{DBLP:journals/tosem/ChenL23a}. For example, it has been shown that for \textsc{Storm}, the default configuration can cause the system $480\times$ slower than the optimal one~\cite{DBLP:conf/mascots/JamshidiC16}.

Configuration tuning is therefore an important task in software engineering, as what have been reported in the literature~\cite{DBLP:journals/tse/SayaghKAP20}. Yet, tuning complex systems is challenging, because:

\begin{figure}[t!]
\centering
\subfloat[Landscape (darker red is better)]{
\includegraphics[width=.48\linewidth]{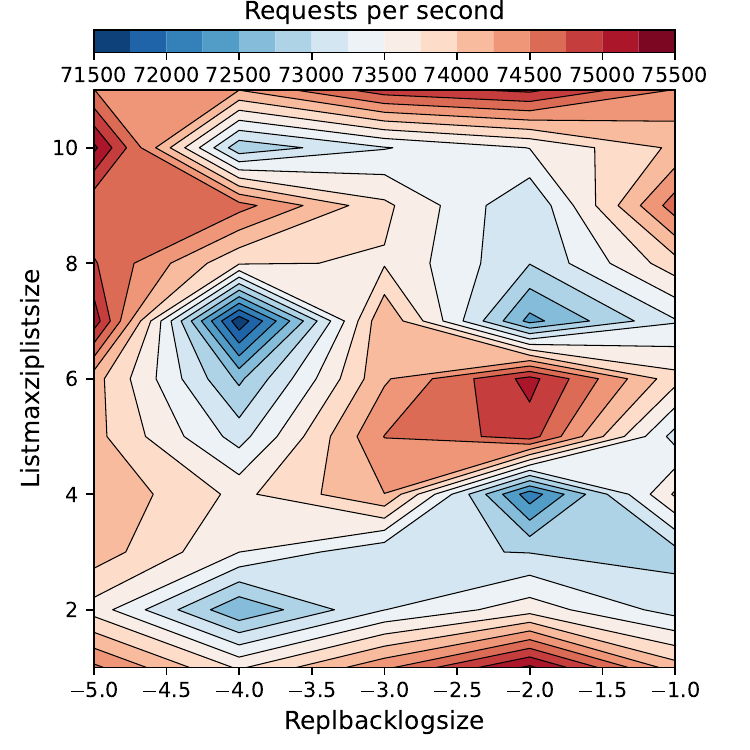}
}
\subfloat[Tuning trajectories]{
\begin{tikzpicture}
    \begin{axis}[
      axis x line  = bottom,
            axis y line  = left  ,
        width=5cm, height=5cm, 
        xlabel={$\#$ of measurements},
        ylabel={Requsts per second},
        ymin=65000,ymax=90000,
        grid=minor,
        legend style={draw=none,at={(1.2,0.02)}, anchor=south east, legend cell align=left}
    ]

    \addplot[
        blue,
        mark=none,
        ultra thick
    ] coordinates {
(0, 72205.75333)
(5, 80221.96333)
(10, 81662.22333)
(15, 81662.22333)
(20, 81662.22333)
(25, 81662.22333)
(30, 82415.30667)
(35, 82415.30667)
(40, 82415.30667)
(45, 82415.30667)
(50, 82415.30667)
(55, 82415.30667)
(60, 82415.30667)
(65, 82415.30667)
(70, 82415.30667)
(75, 82415.30667)
(80, 82415.30667)
(85, 82415.30667)
(90, 82415.30667)
(95, 82415.30667)
(100, 82415.30667)
    };
    \addlegendentry{\texttt{SMAC}}

    \addplot[
        red,
        mark=none,
        ultra thick
    ] coordinates {
(0, 72826.76333)
(5, 75982.76333)
(10, 77765.80667)
(15, 77765.80667)
(20, 80172.72333)
(25, 80172.72333)
(30, 80172.72333)
(35, 80172.72333)
(40, 80172.72333)
(45, 80172.72333)
(50, 80172.72333)
(55, 80172.72333)
(60, 80172.72333)
(65, 80172.72333)
(70, 80172.72333)
(75, 80172.72333)
(80, 80172.72333)
(85, 80319.25667)
(90, 80319.25667)
(95, 80319.25667)
(100, 80319.25667)
    };
     \addlegendentry{Random Search}

    \addplot[
        teal,
        mark=none,
        ultra thick
    ] coordinates {
        (0.0, 88122.34667) (100.0, 88122.34667)
    };
    \addlegendentry{Optimality}

    \end{axis}
\end{tikzpicture}
}

\caption{Example of \textsc{Redis} system. (a) is the projected configuration landscape; (b) is the tuning trajectories of two tuners.}
\label{fig:exp}
\vspace{-0.5cm}
\end{figure}
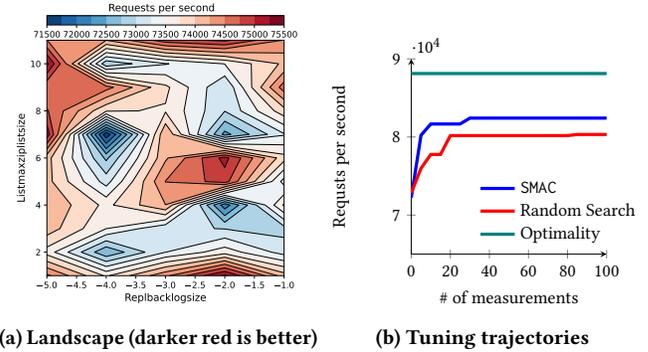

\begin{itemize}
    \item The number of possible configurations can be huge. For example, for the system \textsc{7z}, 14 options have already led to more than a million configurations.
    \item Configuration landscape is highly rugged/sparse~\cite{10832565,DBLP:journals/tosem/GongC25}, meaning that there can be different local optima that might ``trap'' the tuning (see Figure~\ref{fig:exp}a). This makes sense, because if an option is to change the cache strategy, then it would significantly impact the performance. However, in the tuning, it is merely represented as a single-digit change.
    \item The measurement can be extremely expensive~\cite{DBLP:conf/cloud/ZhuLGBMLSY17, DBLP:journals/tse/ChenNKM19, DBLP:conf/sigsoft/0001L24,DBLP:conf/sigsoft/0001L21,DBLP:conf/icse/MaChen25}. For example, it takes more than $1,536$ hours to sample the configurations of 11 options for \textsc{x264}~\cite{DBLP:conf/wosp/ValovPGFC17}. Therefore, tens or hundreds of measurements are common budgets~\cite{DBLP:journals/tse/Nair0MSA20,DBLP:conf/icse/0003XC021}. 
     
\end{itemize}

As such, exhaustively profiling configurable system is unrealistic, thus in the past decade the research community has proposed various tuners based on heuristics~\cite{DBLP:journals/tosem/ChenL23,DBLP:conf/icse/YeChen25,DBLP:journals/tse/ChenNKM19, DBLP:conf/sigsoft/0001L24,DBLP:conf/sigsoft/0001L21,DBLP:conf/sc/BehzadLHBPAKS13,DBLP:conf/sigsoft/ShahbazianKBM20, DBLP:journals/tse/Nair0MSA20,DBLP:conf/kbse/ZhuH23}, which are less sensitivity to the size of configuration space. However, those tuners often need to handle a difficult balance of how to spend the budget: either exploring uncertain regions (for escaping from local optima) or exploiting the known good configurations to guide (for fast convergence)~\cite{DBLP:journals/corr/abs-2112-07303}. The former refers to exploration, meaning that more budget would be consumed for randomly jumping out from local optima under uncertainty, but there is no guarantee that the budget used would bring benefits; while the latter, which focuses on exploitation, uses more budget to search around the good configurations found so far, but it might easily lead to premature convergence at local optima. Because of the above, existing tuners can still struggle to tune certain systems. The fact that those tuners are mostly black-box further exacerbates this issue, as there is no explainability provided on the configuration landscape.



Figure~\ref{fig:exp}b shows an example: we see that both the model-based tuner \texttt{SMAC}~\cite{SMAC} and Random Search struggle, but due to completely opposed causes: \texttt{SMAC} adopts a greedy local search heuristic in the model space, hence it is highly efficient in using the budget to guide the tuning based on good configurations found, but can easily be trapped at local optima with premature convergence. In contrast, Random Search is naturally resilient to local optima, but it lacks strong guidance to efficiently utilize the budget for converging.

In this paper, we take a different perspective on the above limitation and challenges: drawing on the observation that, in general, most of the good configurations tend to be more condensed to certain \textbf{\textit{promising region(s)}} in the configuration space~\cite{DBLP:journals/tse/Nair0MSA20,10832565,DBLP:journals/corr/abs-2112-07303}, we hypothesize that lacking the knowledge of those promising regions can be the root cause of the above ineffective tuning, complicating the issue of balancing budget for exploration and exploitation. To that end, we present \tuner, a tuner guided by causally-verified promising regions with explainability. The key idea is that we learn rules that bound the configuration landscape as the representation of regions and exploit causal inference to causally purify the rules that approximately reflect the promising regions. These rules, which can be iteratively updated and are self-explainable, would then guide a model-based Bayesian optimizer, mitigating the impact of exploration and exploitation trade-offs.

What makes \tuner~unique is that, unlike existing work where causality has been used to analyze configuration options~\cite{DBLP:conf/eurosys/IqbalKJRJ22,DBLP:journals/corr/abs-2402-05399}, we use it to purify the regions in the configuration space, as represented by rules, hence providing finer-grained control over the tuning. The purified rules, after further filtering using all measured configurations by the end of tuning, can be used to better explain the behaviors of the configurable system at a fine-grained landscape level. In a nutshell, our contributions are:

\begin{itemize}
    \item We extract the paths learned by a Random Forest---which is predominantly used in the configuration tuning and handles sparsity well~\cite{DBLP:conf/icse/0003XC021,SMAC}---as the rules and featurize them with the measured configurations, making them causally analyzable.  
    \item Rules are purified by causal relations and effects, identifying those that can approximately reflect the promising regions.
    \item The purified rules guide a model-based Bayesian optimizer while being dually updated with the performance model.
    \item \tuner~extracts the rules that can be fitted by top\% performing configurations, providing explainability on the spatial aspect at the level of configuration landscape.
    \item We assess \tuner~by comparing it with 11 diverse state-of-the-art tuners, including one that leverages causal inference for analyzing options with explainability.
\end{itemize}

The results are encouraging: we reveal that \tuner~performs considerably superior to the state-of-the-art tuners with at least $42\%$ better rank, which is solely contributed by the causally-purified rules. Most importantly, the explainability of \texttt{PromiseTune} at the landscape level can provide richer spatial information that has not been covered in existing option level explainable tuners. All source code and data can be found at our repository: 
\begin{displayquote}
\textcolor{blue}{\texttt{\href{https://github.com/ideas-labo/PromiseTune}{https://github.com/ideas-labo/PromiseTune}}} 
\end{displayquote}

The remainder of the paper is as follows: Section~\ref{sec:preliminaries} presents the preliminaries. Section~\ref{sec:approach} specifies \tuner~designs. Section~\ref{sec:experiments} elaborates on the experiment setup, followed by the results in Section~\ref{sec:results}. Section~\ref{sec:discussion} presents a discussion. Section~\ref{sec:threats},~\ref{sec:related_work}, and~\ref{sec:conclusion} present threats to validity, related work, and conclusion, respectively.

\section{Preliminaries}
\label{sec:preliminaries}

\subsection{Problem Formulation}

In general, the goal of configuration tuning is to optimize a performance metric, e.g., latency or throughput, subject to a budget:
\begin{equation}
		\arg\min f(\vect{c}) \text{ or } \arg\max f(\vect{c})
\end{equation}
where $\vect{c} = \{o_{1}, o_{2}, \dots, o_{n}\}$ is the optimal configuration such that $o_n$ is a configuration option, which can be a binary, integer, or enumerated value. $f$ denotes measuring the system for evaluating the performance obtained by setting a certain configuration.



\begin{figure}[t!]
\centering
\subfloat[\textsc{Redis} (darker red is better)]{
\includegraphics[width=.495\linewidth]{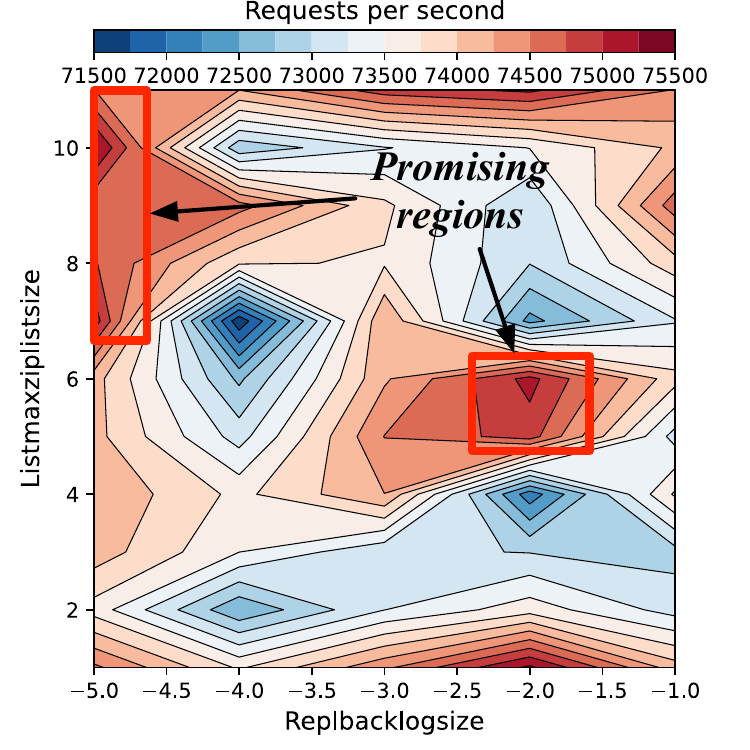}
\label{fig: system-a}
}
\subfloat[\textsc{JavaGC} (darker blue is better)]{
\includegraphics[width=.495\linewidth]{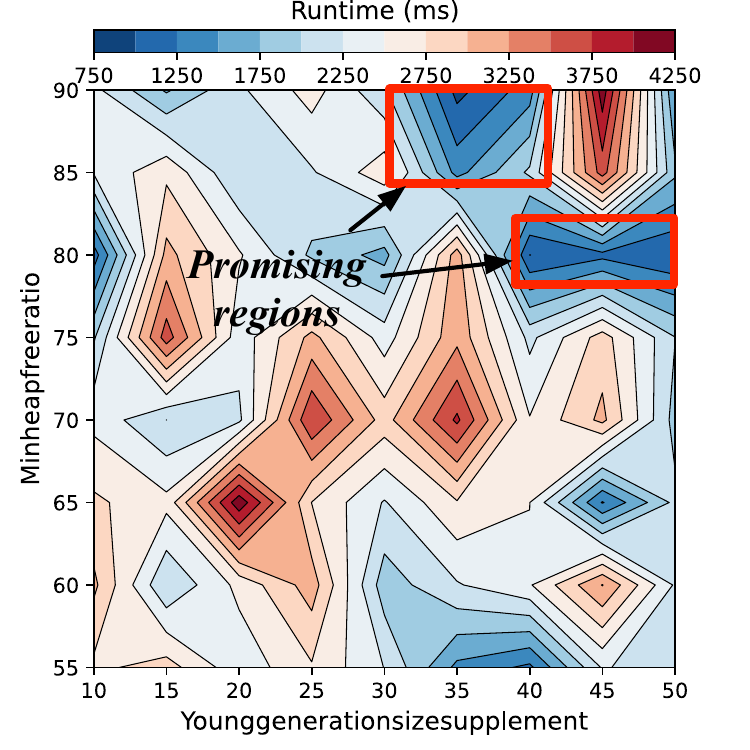}
\label{fig: system-b}
}

\caption{Projection of the configuration landscape for two systems with respect to the performance and two key options.}
\label{fig:system}
\vspace{-0.4cm}
\end{figure}

\subsection{Unaddressed Challenges in Tuning}


Tuning configurations have various known properties, among which the most relevant ones to a heuristic-based tuner are:


\begin{itemize}
    \item Rugged configuration landscape with diverse local optima.
    \item Costly measurements of the configurations.
\end{itemize}

Existing tuners that seek to overcome local optima might consume many resources to explore uncertain regions in the configuration landscape~\cite{DBLP:journals/pvldb/KanellisDKMCV22,DBLP:journals/corr/abs-2112-07303}; while those that tend to exploit most measurements to focus on the best region found so far might stick at local optima forever~\cite{SMAC,DBLP:journals/tse/Nair0MSA20,DBLP:conf/icse/0003XC021}. To understand the root causes, we analyze the landscape of configurable systems. Figure~\ref{fig:system} shows two examples, from which we observe the following spatial information:

\begin{itemize}
    \item Bad and undesired configurations can spread over different regions in the landscape, as can be seen for both systems;
    \item but most good configurations tend to condense in certain \textbf{\textit{promising region(s)}}, e.g., when \texttt{Listmaxziplistsize} $\geq6.7$ and \texttt{Replbacklogsize} $\leq4.7$ or $4.8\leq$ \texttt{Listmaxziplist} \texttt{size} $\leq6.4$ and $-2.4\leq$ \texttt{Replbacklogsize} $\leq-1.6$ for \textsc{Redis}.
\end{itemize}


The above is a corollary of the high ruggedness/sparsity in configuration landscape, which has been discussed in \texttt{FLASH}~\cite{DBLP:journals/tse/Nair0MSA20} (point 5; page 801), and more recently by Chen et al.~\cite{10832565,DBLP:journals/corr/abs-2112-07303}.

The absence of knowledge on the promising region(s) explains the issues in existing tuners: when overcoming local optima (i.e., exploration), the tuning might be forced to jump and explore irrelevant regions, even if it has already reached the promising region(s); when leveraging the neighborhood of the good configurations found (exploitation) to push the tuning, it might get stuck at unwanted local optima if those configurations are far away from the promising region(s). Neither of the above is ideal.

This thus motivates our idea: \textit{what if there is a way to spatially approximate where the promising regions are, and use that to guide the tuner?} As such, we would not only be able to mitigate the impact of exploration and exploitation trade-offs but also spatially explain why certain configurations are better, assisting the designs of tuners and configurable systems. Yet, the challenges are three-fold:

\begin{itemize}
    \item \textbf{Challenge 1:} How to represent/identify promising region(s)?
    \item \textbf{Challenge 2:} How to leverage the promising region(s) in guiding the tuning?
    \item \textbf{Challenge 3:} How to leverage these promising regions to spatially explain the configuration landscape?
\end{itemize}

The above are the key challenges that we address in this paper.

\begin{figure}[t]
    \centering
    \includegraphics[width=\linewidth]{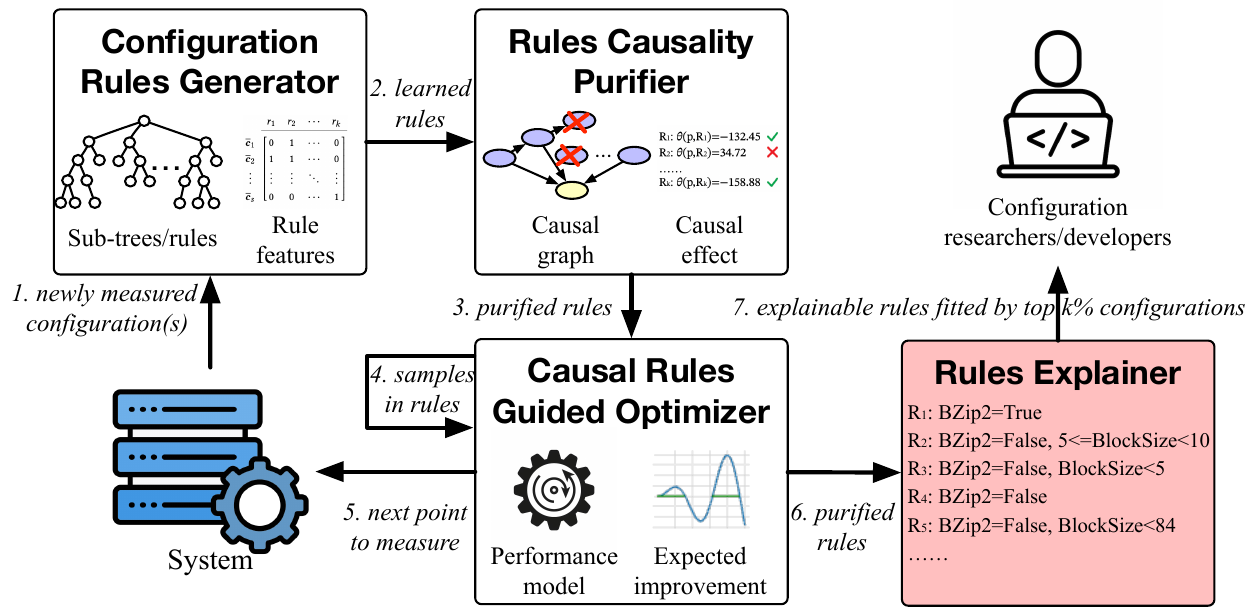}
    \caption{Workflow overview of \tuner.}
    \label{fig:overview}
    \vspace{-0.3cm}
\end{figure}

\section{Tuning with Causally Promising Regions}
\label{sec:approach}
Figure~\ref{fig:overview} shows the workflow of \tuner. Here, the key idea is to leverage configuration rules, learned by Random Forest, to represent the regions in the configuration landscape. Those rules would then be further purified via causal inference, leaving only the rules that reflect the promising regions. As such, the causality is used to analyze the implications of regions (represented as rules) in the configuration landscape as opposed to the impact of options that is commonly used in prior work~\cite{DBLP:conf/eurosys/IqbalKJRJ22,DBLP:journals/corr/abs-2402-05399}. The purified promising regions can then bound and guide a model-based Bayesian optimizer that uses Random Forest as the surrogate/performance model. The rule learning (via the Random Forest) and purification (via causal discovery), together with the performance model, are updated iteratively during tuning, making them incrementally more useful. \tuner~has the following key components:

\begin{figure}[t]
    \centering
    \includegraphics[width=\linewidth]{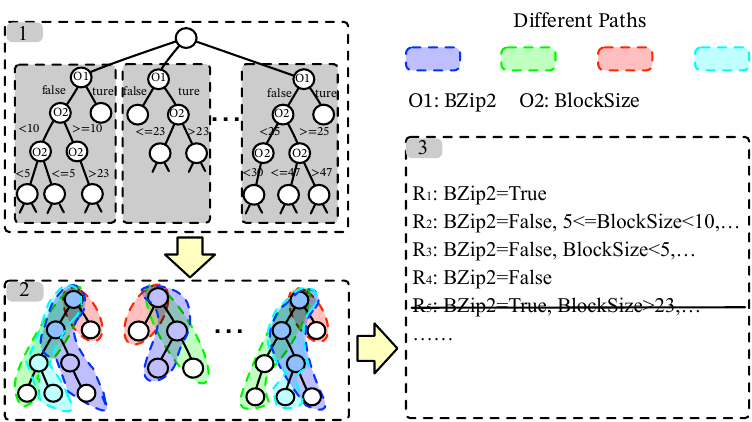}
    \caption{Simplified example of rule learning on \textsc{7z}.}
    \label{fig:rule-gen}
    \vspace{-0.3cm}
\end{figure}

\begin{itemize}
    \item \textbf{Configuration Rules Generator (lines 4--6)} learns rules from measured configuration and featurizes them into a quantifiable format (\textbf{\textit{Challenge 1}}).
    \item \textbf{Rules Causality Purifier (lines 7--11)} purifies the learned rules, identifying those that approximately reflect the promising regions via causal relations and effects (\textbf{\textit{Challenge 1}}).
     \item \textbf{Causal Rules Guided Optimizer (lines 12--20)} is guided by the purified rules to tune configurations (\textbf{\textit{Challenge 2}}).
    \item When tuning terminates, \textbf{Rules Explainer (line 22)} correlates the purified rules with the measured performance, presenting spatially explainable rules fitted by top performance to the researchers/developers (\textbf{\textit{Challenge 3}}).

\end{itemize}

By approximating the promising regions, \tuner~can naturally mitigate the impact of exploration and exploitation trade-offs in the tuning. Detailed procedure can be found in Algorithm~\ref{alg:p-code} and Table~\ref{tb: notations} summarizes the notations used throughout the paper.

\begin{table*}[t!]
\centering
\footnotesize
\setlength{\tabcolsep}{1mm}
\caption{Key notations and their descriptions used in this work.}
\label{tb: notations}

\begin{adjustbox}{width=\linewidth,center}


\begin{tabular}{ll|ll}
\toprule
\textbf{Notation} & \textbf{Description} & \textbf{Notation} & \textbf{Description} \\ \hline

$B$&Tuning budget&$R_i$&The $i$th rule from a set\\

\rowcolor{blue!10}$s$&Initial sample size&$\vect{p}$&Data of the performance metric\\

$l$&Minimal number of leaves for the Random Forest that learns the rules&$\mathbfcal{R}_p$&Set of finally purified rules from $\mathbfcal{R}_m$ using FCI and causal effect\\

\rowcolor{blue!10}$k$&Top $k\%$ measured configurations that verifies the rules for explainability&$\mathbfcal{F}_{perf}$&Random Forest as the surrogate model in Bayesian optimization\\

$\mathbfcal{S}$&A set of configuration samples and their performance values&$\mathbfcal{C'}$&Temporary set of configurations sampled under a rule/region $R_i$\\

\rowcolor{blue!10}$b$&The consumed budget so far&$\mathbfcal{C}$&Set of configurations sampled under all the rules/regions in $\mathbfcal{R}_p$\\

$\mathbfcal{F}_{rule}$&Random Forest that learns the rules&$c'_{best}$&The best configuration on acquisition for the current iteration\\

\rowcolor{blue!10}$\mathbfcal{R}_{l}$&Rules extracted from Random Forest $\mathbfcal{F}_{rule}$&$c_{best}$&The best configuration on acquisition for all iterations\\

$\mathbfcal{S}'$&Samples of configurations featurized/represented by rules via fitting $\mathbfcal{R}_{l}$ and $\mathbfcal{S}$&$\mathbfcal{R}'_p$&Set of explainable rules from $\mathbfcal{R}_p$ after verifying with the top $k\%$\\

\cellcolor{blue!10}$\mathbfcal{R}_m$&\cellcolor{blue!10}Set of intermediate rules purified using FCI only via $\mathbfcal{S}'$&&sampled configurations\\

\bottomrule
\end{tabular}

\end{adjustbox}

\end{table*}


\begin{algorithm}[t!]
	\DontPrintSemicolon
	
	\caption{Pseudo code of \tuner}
	\label{alg:p-code}
	\KwIn{Budget $B$; initial sample size $s$; parameters $l$ and $k$}
  \KwOut{The best configuration found $\vect{c}_{best}$; the extracted rules for explainability $\mathbfcal{R}'_p$}
    
    $\mathbfcal{S}\leftarrow$ measure $s$ configurations via random sampling; each sample is a configuration-performance pair (Equation~\ref{eq:matrix1})\\
     \For{$b + s < B$}  
      {
         reset $\mathbfcal{F}_{rule}$, $\mathbfcal{F}_{perf}$, $\mathbfcal{S'}$, $\mathbfcal{C}$, $\mathbfcal{R}_l$, $\mathbfcal{R}_m$, $\mathbfcal{R}_p$ as $\emptyset$\\
         $\mathbfcal{F}_{rule}\leftarrow$ train/update a Random Forest using $\mathbfcal{S}$ with $l$\\
         $\mathbfcal{R}_l\leftarrow$ learn and extract rules from $\mathbfcal{F}_{rule}$\\
         $\mathbfcal{S}'\leftarrow$ featurize $\mathbfcal{R}_l$ into $\mathbfcal{S}$ as Equations~\ref{eq:matrix1} and~\ref{eq:matrix2}\\
         $\mathbfcal{R}_m\leftarrow$ purify $\mathbfcal{R}_l$ via the FCI-built causal graph over $\mathbfcal{S}'$\\
          \For{$\forall R_i \in \mathbfcal{R}_m$}  
          {
             $\theta(p,R_i)\leftarrow$ compute via Equation~\ref{ACE}\\
             \If{$\theta(p,R_i)$ < 0} {
             $\mathbfcal{R}_p\leftarrow\mathbfcal{R}_p\cup R_i$
             }
          }
          $\mathbfcal{F}_{perf}\leftarrow$ train/update a Random Forest using $\mathbfcal{S}$\\

            \For{$\forall R_i \in \mathbfcal{R}_p$}  
          {

              \While{sample more for $\mathbfcal{C}'$ can still improve $\alpha_{EI}$}  
              {
                  $\mathbfcal{C}'\leftarrow$ randomly sample a configuration from the region bounded by $R_i$ and evaluate it via $\mathbfcal{F}_{perf}$ and Equation~\ref{EI}\\
              }
              $\mathbfcal{C}\leftarrow\mathbfcal{C}\cup\mathbfcal{C}'$\\
              reset $\mathbfcal{C}'=\emptyset$\\

          }
          $\{\vect{c}'_{best},p\}\leftarrow$ get the configuration from $\mathbfcal{C}$ with the best $\alpha_{EI}$ and measure it on the system for its performance $p$\\
          $\mathbfcal{S}\leftarrow\mathbfcal{S}\cup\{\vect{c}'_{best},p\}$\\
          $b=b+1$\\
          
      }

      $\vect{c}_{best}=$ the configuration from $\mathbfcal{S}$ with the best performance\\
      $\mathbfcal{R}'_p\leftarrow$ extract explainable rules from $\mathbfcal{R}_p$ that fit the top $k$\% performing configurations from $\mathbfcal{S}$\\  
      \Return $\{\vect{c}_{best},\mathbfcal{R}'_p\}$

\end{algorithm}

\subsection{Configuration Rules Generation}
\label{sec:rule-gen}

\subsubsection{Learning Rules}

Given a set of measured configurations $\mathbfcal{S}$, \tuner~leverages Random Forest~\cite{breiman2001random}, denoted as $\mathbfcal{F}_{rule}$, to learn and represent the regions (a common and pragmatic choice). Random Forest builds a set of sub-trees, each of which consists of different paths that partially traverse the configuration space\footnote{Note that an option in the sub-tree, which is a node, can be further split. For example, if \texttt{BlockSize} $>10$ is a path from one split, then the next split paths can still be \texttt{BlockSize} $\leq15$ and \texttt{BlockSize} $>15$.}. Each of the paths forms a \textit{\textbf{learned rule}}, bounding a region in the configuration landscape. As in Figure~\ref{fig:rule-gen}, we perform the following:

\begin{enumerate}
    \item Train a Random Forest to correlate configurations and their measured performance using sample set $\mathbfcal{S}$.
    \item Extract every path from the sub-trees as a rule, which not only eliminates trivial options but also bounds the landscape.
    \item Merge the overlapping ranges of an option in a rule using their intersection and remove duplicated rules. For example, the rule $\langle$\texttt{BZip2=True}, \texttt{BlockSize} $<7$, \texttt{BlockSize} $<5\rangle$ for \textsc{7z} can be merged as $\langle$\texttt{BZip2=True}, \texttt{BlockSize}$<5\rangle$.
\end{enumerate}

This process will produce a set of unique rules, such as $\langle$\texttt{BZip2=True}, $5\leq$ \texttt{BlockSize} $<10\rangle$ for \textsc{7z}. It is possible that the region bounded by a rule is a partial or full subset of that bounded by the other, implying that the overlapped parts are important (see Section~\ref{sec:exp-thoery}).

The Random Forest has a key parameter $l$ that controls the minimal number of leaves, which is important for \tuner~as it directly determines the minimal number of paths in the sub-trees, and hence the smallest number of rules learned. This can influence both the performance and explainability of \tuner. In Section~\ref{sec:rq3}, we will study the sensitivity of \tuner~to $l$.

\subsubsection{Featurizing Rules}

Although the rules are useful representations of the regions in configuration landscapes, we still need to link them to the sampled configurations' performance for further quantification and analysis. To that end, \tuner~``featurizes'' the rules by converting them into the features for the configurations. 


Recall that given a configuration $\vect{c} = \{o_{1}, o_{2}, \dots, o_{n}\}$ and a set of learned rules $\mathbfcal{R}_l = \{{R}_{1}, {R}_{2}, \dots, {R}_{k}\}$, we represent the configuration as $\vect{c} = \{r_{1}, r_{2}, \dots, r_{k}\}$ where $r_k$ is a binary feature/value that indicates whether the configuration $\vect{c}$ fits the $k$th rule:

\begin{itemize}
    \item A configuration \textbf{fits} the rule if it fails within the region bounded by the rule, i.e., the values of the configuration meet with all the bounded options in the rule\footnote{For (unbounded) options not covered by a rule, any permitted values are allowed.} ($r_k=1$).
    \item Otherwise, any violation of a configuration's value over an option covered by the rule would make it \textbf{violated} ($r_k=0$).
\end{itemize}


For example, if there are two rules ${R}_{1}=\langle$\texttt{BZip2=True}$\rangle$ and ${R}_{2}=\langle$\texttt{BZip2=False}, $5\leq$ \texttt{BlockSize} $<10\rangle$, along with a configuration $\vect{c}$ originally as $\{0,8\}$ (for binary options, $1$ denotes True or $0$ otherwise), then $\vect{c}$ fits ${R}_{2}$ (${r}_2=1$) but not ${R}_{1}$ (${r}_1=0$), hence the configuration becomes $\{0,1\}$ after featurizing with the rules.


We featurizing the rules over all configurations, transforming into a newly customized dataset, e.g., suppose that we have a set ($\mathbfcal{S}$) of $s$ configurations with their measured performance $\vect{p}$: 
\begin{equation}
\label{eq:matrix1}
 \begin{blockarray}{c cccccc}
& o_1  & o_2 & \cdots & o_n & & \vect{p} \text{ runtime (ms)}  \\
\cmidrule{2-5}\cmidrule{7-7}
\begin{block}{c [cccc]c[c]}
\vect{c}_1 &  0 & 8 & \cdots & 9 & = & 22057.7  \\[-0.6ex]
\vect{c}_2 &  1 & 15 & \cdots & 2 & = & 12300.3  \\[-0.6ex]
\vdots & \vdots & \vdots & \ddots & \vdots & \vdots & \vdots  \\[-0.6ex]
\vect{c}_s &  1 & 3 & \cdots & 0 & = & 55320.6 \\
\end{block}
\noalign{\vspace{-100.5ex}}
\\
\end{blockarray}
\end{equation}
After featurizing the rules, we obtain a new dataset as:
\begin{equation}
\label{eq:matrix2}
 \begin{blockarray}{c cccccc}
& r_1  & r_2 & \cdots & r_k & & \vect{p} \text{ runtime (ms)}  \\
\cmidrule{2-5}\cmidrule{7-7}
\begin{block}{c [cccc]c[c]}
\vect{c}_1 &  0 & 1 & \cdots & 0 & = & 22057.7 \\[-0.6ex]
\vect{c}_2 &  1 & 0 & \cdots & 0 & = & 12300.3 \\[-0.6ex]
\vdots & \vdots & \vdots & \ddots & \vdots & \vdots & \vdots   \\[-0.6ex]
\vect{c}_s &  0 & 0 & \cdots & 1 & = & 55320.6  \\
\end{block}
\noalign{\vspace{-100.5ex}}
\\
\end{blockarray}
\end{equation}

Here, the dimensions might vary as $n$ and $k$ could differ\footnote{A configuration might fit more than one rule.}. The causality of the new dataset will later on be analyzed.

\subsection{Causally Purifying Configuration Rules}
\label{sec:pur-rules}

\begin{figure}[t]
    \centering
    \subfloat[Causal graph]{
\includegraphics[width=.61\linewidth]{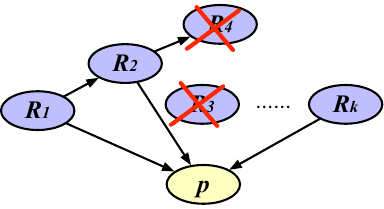}
}
\subfloat[Causal effects]{
\includegraphics[width=.37\linewidth]{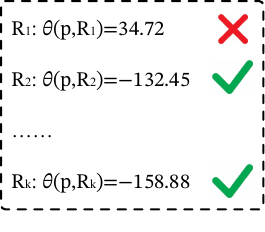}
}
    \caption{Example of purification via causal inference.}
    \label{fig:causal}
    \vspace{-0.2cm}
\end{figure}

\subsubsection{Purifying via Causal Graph}

With the newly obtained configuration representation, it is easy to pair each configuration with its measured performance. Yet, not all the learned rules can reflect the promising regions, hence a purification is needed. To that end, we then feed the entire new dataset into Fast Causal Inference (FCI) \cite{DBLP:conf/uai/SpirtesMR95}---a causal discovery algorithm---for analyzing the causal relations between the configurations represented by rules (as features) and the performance, together with those between rules, because: 


\begin{itemize}
    \item FCI works better than the others, e.g., the Peter-Clark algorithm~\cite{spirtes2001causation}, on handling unobserved confounders. 
    \item FCI makes fewer assumptions on data, e.g., algorithms like LinGAM~\cite{shimizu2014lingam} require linear causal relations.
\end{itemize}

In a nutshell, FCI builds a complete undirected graph on all rules and the performance metric, from which the edges are removed if two rules (or a rule and the performance) are conditionally independent. FCI also orients the edges using collider detection and according to latent confounders, leading to a partial ancestral graph.

With the graph produced by FCI, \tuner~then eliminates those rules that are not involved in any path that ends at the performance as they are unlikely to reflect the promising regions, creating an intermediate rule set $\mathbfcal{R}_m$. Figure~\ref{fig:causal}a shows an example where all vertices and edges are produced by FCI; the arrows indicate causal relations; crosses highlight the rules eliminated by \tuner, since $R_3$ and $R_4$ are not part of any paths that end at $\vect{p}$.

\subsubsection{Purifying via Causal Effects}

As in Figure~\ref{fig:causal}b, drawing on the causal graph, \tuner~computes the average causal effect for a rule $R_i$ ($R_i \in \mathbfcal{R}_m$) on the performance $\vect{p}$ in do-calculus~\cite{pearl2000models} as:
\begin{equation}
    \theta(\vect{p},R_i)= \mathbb{E}[f|do(r_i=1)]-\mathbb{E}[f|do(r_i=0)]
    \label{ACE} 
\end{equation}
whereby $\mathbb{E}[f|do(r_i=1)]$ and $\mathbb{E}[f|do(r_i=0)]$ are the expected performance change for all configurations that fit and violate $R_i$, respectively, as computed by FCI. We can easily find the fitted and violated configurations by examining the transformed dataset with rule features in Equation~\ref{eq:matrix2}, i.e., for $R_i$, those configurations with $r_i=1$ are the fitted ones, or otherwise they are violated if $r_i=0$. $\theta$ can be positive or negative, but a smaller value is preferred for minimized performance metrics. \tuner~further purifies the rules by discarding those with $\theta\geq0$ as this indicates that when configurations fit them, the performance can actually be worsened (or no change). The remaining rules with $\theta<0$, denoted as $\mathbfcal{R}_p$, are the \textit{\textbf{purified rules}} that serve as good approximations of the promising regions. $\mathbfcal{R}_p$ can then be used to guide the tuning and explain the configuration landscape. 

For example, under a sample size of $50$, ${R}_{2}=\langle$\texttt{BZip2=False}, $5\leq$ \texttt{BlockSize} $<10\rangle$ can have $27$ fitted configurations and $23$ violated ones, leading to $\mathbb{E}$ of $354.44$ and $486.89$, respectively, and hence we have $\theta=-132.45$. ${R}_{2}$ should therefore be included in $\mathbfcal{R}_p$. Note that the number of fitted and violated configurations for a rule are often of similar quantity, because those insignificant rules commonly have limited causal relationships to the performance, and hence should have been removed as part of the causal graph purification.

Noteworthily, since configuration tuning does not often have a large amount of data to mine highly accurate causal relations, here we adopt a coarse-grained strategy rather than a fine-grained one: we are interested in whether the rule can improve performance or worsen it, rather than the extent of such improvement/degradation.

\subsection{Causal Rules Guided Optimization}

The rules with $\theta<0$ provide insights into the approximated promising regions. As a result, \tuner~leverage this information in the exploration process of the tuning. While theoretically, those rules can benefit different optimizers, we found that they are particularly useful when paired with a variant of the model-based Bayesian optimizer that leverages Random Forest ($\mathbfcal{F}_{perf}$) as the surrogate/performance model. Assuming minimization, we use Expected Improvement (EI)~\cite{zhan2020expected} as the acquisition function:
\begin{equation}
    \alpha_{EI}(\vect{c})= \mathbb{E}_{f(\vect{c})}\max(0,f(\vect{c}) - p_{best})
     \label{EI} 
\end{equation}
whereby $\alpha_{EI}(\vect{c})$ is the EI value of $\vect{c}$; $f(\vect{c})$ is the performance of $\vect{c}$ predicted by the performance model; $p_{best}$ is the best (predicted) performance observed so far.

Specifically, \tuner~uses the promising regions represented as causally purified rules in the steps below to guide the tuning:

\begin{table*}[t!]
\centering
\footnotesize
\caption{Details of the subject systems with diverse domains, performance metrics to be optimized, and sizes of configuration/search space $\mathbfcal{S}_{space}$. ($|\mathbfcal{B}|$/$|\mathbfcal{N}|$) denotes the number of binary/numeric options.}
\label{tb: systems}

\begin{adjustbox}{width=\linewidth,center}


\begin{tabular}{lllllllll}
\toprule
\textbf{System} & \textbf{Version} & \textbf{Benchmark} & \textbf{Domain} & \textbf{Language} & \textbf{Performance to be optimized} & \textbf{$|\mathbfcal{B}|$/$|\mathbfcal{N}|$}  & \textbf{$\mathbfcal{S}_{space}$} & \textbf{Used by} \\ \hline

\textsc{7z} & 9.20 & Compressing a 3 GB directory & File Compressor & C$++$ & Runtime (ms) & 11/3  & $1.68 \times 10^8$ & \cite{DBLP:conf/icse/WeberKSAS23} \\
\rowcolor{blue!10}\textsc{DConvert} & 1.0.0 & Transform resources at different scales & Image Scaling & Java & Runtime (s) & 17/1 &  $1.05 \times 10^7$ & \cite{10172849} \\
\textsc{ExaStencils} & 1.2 & Default benchmarks & Code Generator & Scala & Runtime (ms) & 7/5  & $1.61 \times 10^9$ & \cite{DBLP:conf/icse/WeberKSAS23} \\ 
\rowcolor{blue!10}\textsc{BDB-C} & 18.0 & Benchmark provided by vendor  & Database & C & Latency (s) & 16/0  & $6.55 \times 10^4$ & \cite{10832565} \\ 
\textsc{DeepArch} & 2.2.4 & UCR Archive time series dataset & Deep Learning Tool & Python & Runtime (min) & 12/0  & $4.10 \times 10^3$ & \cite{DBLP:conf/sigsoft/JamshidiVKS18} \\ 
\rowcolor{blue!10}\textsc{PostgreSQL} & 22.0 & PolePosition 0.6.0 & Database & C & Runtime (ms) & 6/3  & $1.42 \times 10^9$ & \cite{DBLP:conf/icse/WeberKSAS23} \\ 
\textsc{JavaGC} & 7.0 & DaCapo benchmark suite & Java Runtime & Java & Runtime (ms) & 12/23  & $2.67 \times 10^{41}$ & \cite{10832565} \\ 
\rowcolor{blue!10}\textsc{Storm} & 0.9.5 & Randomly generated benchmark & Data Analytics & Clojure & Messages per Second & 12/0  & $4.10 \times 10^{3}$ & \cite{DBLP:journals/tse/KrishnaNJM21} \\ 
\textsc{x264} & 0.157 & Video files of various sizes & Video Encoder & C & Peak signal-to-noise ratio & 4/13  & $6.43 \times 10^{26}$ & \cite{DBLP:journals/tse/KrishnaNJM21}  \\ 
\rowcolor{blue!10}\textsc{Redis} & 6.0 & Sysbench & Database & C & Requests per second & 1/8  & $5.78 \times 10^{16}$ & \cite{DBLP:journals/jss/CaoBWZLZ23} \\ 
\textsc{HSQLDB} & 19.0 & PolePosition 0.6.0 & Database & Java & Runtime (ms) & 18/0  & $2.62 \times 10^5$ & \cite{DBLP:conf/icse/WeberKSAS23} \\ 
\rowcolor{blue!10}\textsc{LLVM} & 3.0 & LLVM’s test suite & Compiler & C$++$ & Runtime (ms) & 10/0  & $1.02 \times 10^3$ & \cite{10832565} \\ 

\bottomrule
\end{tabular}

\end{adjustbox}

\end{table*}

\begin{enumerate}
    \item Measure initial configuration data using random sampling.
    \item Learn and purify a rule set $\mathbfcal{R}_p$ as stated in Sections~\ref{sec:rule-gen} and~\ref{sec:pur-rules}, and update the performance model $\mathbfcal{F}_{perf}$.
    \item Pick a rule $R_i$ from $\mathbfcal{R}_p$.
    \item Randomly sample configurations in the region\footnote{Note that for options absent from the rule, we perform random sampling on all values.} bounded by $R_i$ while evaluating them via the performance model and $\alpha_{EI}(\vect{c})$, store them in the sampled set $\mathbfcal{C}$. This sampling can be easily done in the perturbation, e.g., if we need to sample configurations for a rule that covers two options $\langle$\texttt{BZip2=False}, $5\leq$ \texttt{BlockSize} $<10\rangle$, then when perturbing, \tuner~simply only allow their values to be randomly set as $\texttt{BZip2=False}$ and $\texttt{BlockSize} \in [5,10)$, while the other uncovered options can have any permitted values. 
    \item To determine when to stop sampling for $R_i$, we use Gaussian Kernel Density Estimation (GKDE)~\cite{terrell1992variable}. In a nutshell, GKDE serves as a termination predictor for the region under each rule, preventing unnecessary sampling when further samples cannot significantly improve the results. As such, it is complementary to the performance model $\mathbfcal{F}_{perf}$.
    

    \item Repeat from (3) until all rules in $\mathbfcal{R}_p$ have been sampled.
    \item Select the configuration with the best $\alpha_{EI}(\vect{c})$ from $\mathbfcal{C}$ and measure it on the system.
    \item If the budget has not been exhausted, repeat from (2); otherwise, terminate the tuning.
\end{enumerate}


In this way, the exploration in \tuner~is guided by the purified rules, which bound on the approximated promising regions, hence consolidating the tuning quality. Notably, simple/short rules would provide loose guidance while complex/long rules can lead to more constrained tuning direction, both of which are relevant to the parameter $l$, which we will discuss in Section~\ref{sec:rq3}.

\subsection{Explainability with Purified Rules}
\label{sec:exp-thoery}

Instead of simply using all purified rules in $\mathbfcal{R}_p$ and presenting them to the researchers/developers, \tuner~assists in the explainability of promising regions by further extracting those that have indeed led to excellent performance. To that end, by the end of the tuning, we use the measured configurations with top $k$\% performance and examine which are the purified rules that those configurations fit. The ones that can be fitted, referred to as \textit{\textbf{explainable rules}}, are then returned. Both $l$ and $k$ can impact the number of explainable rules, in which $l$ also affects the number of learned and purified rules. While $l$ affects both performance and explainability (Section~\ref{sec:rq3}), $k$ only concerns explainability and is case-dependent (Section~\ref{sec:rq4}): lower $k$ might leave too few explainable rules for analysis, but higher $k$ can cause cognitive fragility on too many explainable rules.



Suppose that for the system \textsc{7z} with 14 options, if there are three explainable rules from \tuner~under $k=10$: ${R}_{1}=\langle$\texttt{BZip2=True},  \texttt{BlockSize} $\geq10$, \texttt{mtOff=False}$\rangle$; ${R}_{2}=\langle$\texttt{BZip2=False}, $5\leq$ BlockSize $<10$, \texttt{mtOff=False}$\rangle$; ${R}_{3}=\langle$\texttt{BlockSize} $\geq20$, \texttt{mtOff=} \texttt{True}$\rangle$, we can make the following explanation on the promising configurations with rich spatial information:

\begin{itemize}
    \item \textbf{Important Options:} Those absent options are unlikely to be helpful/important in finding good configurations.
    \item \textbf{Option Interactions:} If there are two or more rules where $q$ options have different ranges/values but the ranges/values of other options are either all the same or all absent, then those $q$ options are likely to have interaction that would lead to promising configurations. In this way, we can then examine what interactions (and their ranges) more commonly lead to promising configurations. For example, the interaction between \texttt{BZip2} and \texttt{BlockSize} is more important for finding good configurations than that between other pairs, since it can be observed from more rules (i.e., $R_1$ and $R_2$).

    \item \textbf{Promising Regions:} The most common overlapping(s) covered by the most rules above (the absent options are unbounded) is a natural reflection of the most promising regions for the system's configuration landscape.
\end{itemize}


Different stakeholders can benefit from the spatial explainability: the above does not only help researchers on future system-specific tuner design but can also inform developers on how to refactor the system---the latter point means that \textit{while most work focuses on designing a better tuners on a fixed problem, for the first time, \tuner~provides hints on how to change/design the problem (system) such that it can make the system easier to be tuned by a tuner.} These will be further discussed in Sections~\ref{sec:rq4} and~\ref{sec:exp-why}.

\section{Experiment Setup}
\label{sec:experiments}

To evaluate \tuner, we ask four research questions (RQs):
\begin{itemize}
\item \textbf{RQ1:} How does \tuner~perform compared with the state-of-the-art tuners?
\item \textbf{RQ2:} How do the causally purified rules help \tuner?
\item \textbf{RQ3:} What is the sensitivity of \tuner~to $l$?
\item \textbf{RQ4:} How well can \tuner~explain the configuration performance against existing explainable approaches?
\end{itemize}

\textbf{RQ1} evaluates the effectiveness of \tuner~against 
others while \textbf{RQ2} verifies the contribution of causally purified rules. \textbf{RQ3} performs sensitivity analysis of \tuner's key parameters and \textbf{RQ4} examines the usefulness of the resulted explainable rules.

All the experiments are conducted on a high-performance server with Ubuntu 20.04.1 LTS, Intel(R) Xeon(R) Platinum 8480$+$ with 224 CPU cores and 500GB memory.



\subsection{Subject Configurable Systems}

As in Table~\ref{tb: systems}, we examine all the systems and their datasets that have been studied while filtering them based on the following:


\begin{itemize}
    \item For systems of the same domain, language, and performance metric from prior work, we use the one with the highest number of options to tune, e.g., \textsc{BDB-C} and \textsc{MariaDB} are both database systems concerning latency and are written primarily in C, but only \textsc{BDB-C} is used as it often has more options. The same applies to various versions of the same system, e.g., \textsc{Storm} has been studied in many prior
studies~\cite{DBLP:journals/tse/Nair0MSA20,DBLP:journals/corr/abs-2112-07303}, and we use the most complicated case of 12 options.
\item We filter those systems that have no commonly agreed benchmark in different prior studies.
\end{itemize}

The final set consists of 12 systems of diverse domains, options/types, size of configuration space, and languages, e.g., Clojure, C and Java. Therefore, these serve as a comprehensive set of subject systems for evaluation. 

For the options and performance benchmark, we directly use what has been adopted in prior work (see Table~\ref{tb: systems}, rightmost column), focusing only on the performance-sensitive ones~\cite{DBLP:conf/icse/LiangChen25}. 




\subsection{State-of-the-Art Tuners} 
\label{sec:sota}

We compare \tuner~against a wide category of tuners:

\begin{itemize}
    \item \textbf{General:} We use Random Search, \texttt{SMAC}~\cite{DBLP:conf/lion/HutterHL11}, \texttt{GA}~\cite{DBLP:conf/sigsoft/ShahbazianKBM20}, \texttt{MBO}~\cite{DBLP:conf/kdd/LiSZCJLJG0Y0021}, and \texttt{HEBO}~\cite{cowen2022hebo} as the general tuners, as they are common for black-box problems, including configuration tuning~\cite{DBLP:conf/sc/BehzadLHBPAKS13,DBLP:conf/sigsoft/ShahbazianKBM20}.
    \item \textbf{Configuration:} This contains \texttt{FLASH}~\cite{DBLP:journals/tse/Nair0MSA20} and \texttt{Unicorn}~\cite{DBLP:conf/eurosys/IqbalKJRJ22}, both are tuners for general configurable systems from the software engineering community. \texttt{Unicorn} also uses causal inference for explainability, but only at the options level.
    \item \textbf{Compiler:} We pick compiler tuners, i.e., \texttt{BOCA} \cite{DBLP:conf/icse/0003XC021} and \texttt{CFSCA} \cite{DBLP:conf/kbse/ZhuH23}, which are applicable to other configurable systems.
    \item \textbf{Database:} Similarly, we examine the widely used tuners for database systems (\texttt{OtterTune} \cite{DBLP:conf/sigmod/AkenPGZ17} and \texttt{LlamaTune} \cite{DBLP:journals/pvldb/KanellisDKMCV22}), which is one of the most complex systems to tune.
\end{itemize}


The above represents a diverse set of state-of-the-art tuners from different domains and levels of focus. Note that we omit the multi-fidelity tuners such as \texttt{DEHB}~\cite{DBLP:conf/ijcai/AwadMH21}, because although the fidelity for AutoML is well-defined, its definition for general configurable systems is unclear: in AutoML, there exists a fidelity-factor with clear monotonic relationships to the performance metric/cost, which those multi-fidelity tuners have leveraged, e.g., using more training data will have higher-fidelity accuracy but be more costly. For configurable systems, there are no such clear relationships, e.g., on an image rescaling system, it is unclear how the images can be changed to monotonically influence the system performance/cost. As such, comparing with multi-fidelity tuners like \texttt{DEHB} require significant changes, e.g., \texttt{DEHB} would become simply a \texttt{DE}.

\subsection{Budget and Parameter Settings} 

Since the configuration measurement is the most expensive part of configuration tuning~\cite{DBLP:journals/corr/abs-2112-07303,DBLP:journals/tse/Nair0MSA20}, we place budget explicitly on such. To quantify the budget of the tuning, i.e., $B$, we leverage the number of measurements on the real system performance---a widely adopted standard~\cite{DBLP:conf/sigmod/ZhangWCJT0Z021, DBLP:conf/icml/ZiomekB23, DBLP:journals/pvldb/LaoWLWZCCTW24, DBLP:conf/icse/0003XC021, DBLP:conf/sigmod/AkenPGZ17, DBLP:journals/tse/Nair0MSA20}, because it is language- and hardware-independent. Since the measurement of systems is costly, to ensure generality, we test three budget settings: $B \in \{50,100,150,200\}$, where $B=50$ is the smallest considered in the compared tuners (i.e., \texttt{FLASH}). As with prior work, redundant configurations found do not consume the budget~\cite{DBLP:journals/corr/abs-2112-07303,DBLP:journals/tse/Nair0MSA20,DBLP:conf/sigsoft/0001L24}.

To initialize \tuner~and the other model-based tuners (e.g., \texttt{FLASH}), we set an initial sample size of $10$, which is also commonly used~\cite{DBLP:conf/sigmod/AkenPGZ17,DBLP:journals/pvldb/KanellisDKMCV22}. For other parameters, such as the population size of GA, we use the default or set to what has been used in the literature. For \tuner, unless otherwise stated, we set $l=10$ as the most reasonable value, which we will analyze in Section~\ref{sec:rq3}. The $k$ value depends on the explainability scenario (see Section~\ref{sec:rq4}).

We repeat all experiments 30 runs with different seeds. All performance metrics are converted to minimization for better exposition.

\subsection{Statistical Test} 
 


\begin{table*}[t!]


\caption{Comparing \tuner~with state-of-the-art tuners on ``[Scott-Knott ESD rank] mean (standard deviation)'' of the optimized (normalized) performance over 30 runs (the smaller, the better). \colorbox{blue!20}{blue cells} and \colorbox{teal!20}{green cells} denote the tuner(s) with the best and second best Scott-Knott ESD rank for a case, respectively. \textcolor{red}{\ding{55}} denotes incompletion (calculated as the worst rank and 1.0 in overall average). Raw data can be accessed at \textcolor{blue}{\texttt{\href{https://github.com/ideas-labo/PromiseTune/blob/main/RQs/RQ1/rq1.pdf}{https://github.com/ideas-labo/PromiseTune/blob/main/RQs/RQ1/rq1.pdf}}}.}
\vspace{-0.2cm}
\setlength{\tabcolsep}{1mm}
\label{tab: performance}
\begin{adjustbox}{width=\linewidth, center}
  \centering
  \begin{tabular}{llllllllllllll}
    \toprule
   \textbf{Budget} &\textbf{System} & \textbf{\texttt{\tuner}} & \textbf{\texttt{Random}} & \textbf{\texttt{Unicorn}} & \textbf{\texttt{GA}}  & \textbf{\texttt{MBO}} & \textbf{\texttt{LlamaTune}} & \textbf{\texttt{FLASH}} & \textbf{\texttt{CFSCA}} & \textbf{\texttt{BOCA}} & \textbf{\texttt{OtterTune}} & \textbf{\texttt{SMAC}} & \textbf{\texttt{HEBO}}\\
    \midrule
\multirow{14}{*}{$B=50$}

&\textsc{\textbf{7z}}&
[1] \cellcolor{blue!20}0.070 (0.164) &
[2] \cellcolor{teal!20}0.150 (0.218) &
[3] 0.163 (0.217) &
[4] 0.326 (0.301) &
[5] 0.691 (0.002) &
[3] 0.323 (0.338) &
[3] 0.236 (0.265) &
[1] \cellcolor{blue!20}0.120 (0.217) &
[3] 0.241 (0.290) &
[3] 0.293 (0.321) &
[3] 0.188 (0.257) &
[6] 0.990 (0.054) \\
&\textsc{\textbf{DConvert}}&
[2] \cellcolor{teal!20}0.077 (0.072) &
[7] 0.222 (0.121) &
[7] 0.230 (0.151) &
[7] 0.358 (0.320) &
[8] 0.726 (0.082) &
[7] 0.296 (0.272) &
[3] 0.105 (0.084) &
[4] 0.110 (0.150) &
[1] \cellcolor{blue!20}0.070 (0.071) &
[5] 0.115 (0.237) &
[6] 0.176 (0.263) &
[1] \cellcolor{blue!20}0.044 (0.058) \\
&\textsc{\textbf{ExaStencils}}&
[1] \cellcolor{blue!20}0.088 (0.079) &
[3] 0.130 (0.061) &
[3] 0.130 (0.061) &
[4] 0.151 (0.096) &
[5] 0.766 (0.141) &
[2] \cellcolor{teal!20}0.115 (0.067) &
[1] \cellcolor{blue!20}0.079 (0.090) &
[1] \cellcolor{blue!20}0.080 (0.078) &
[2] \cellcolor{teal!20}0.114 (0.078) &
[3] 0.126 (0.078) &
[1] \cellcolor{blue!20}0.092 (0.080) &
[1] \cellcolor{blue!20}0.082 (0.070) \\
&\textsc{\textbf{BDB-C}}&
[3] 0.054 (0.108) &
[3] 0.041 (0.033) &
[2] \cellcolor{teal!20}0.033 (0.019) &
[6] 0.189 (0.257) &
[2] \cellcolor{teal!20}0.020 (0.029) &
[4] 0.087 (0.125) &
[6] 0.243 (0.169) &
[5] 0.105 (0.146) &
[6] 0.192 (0.244) &
[3] 0.035 (0.032) &
[5] 0.126 (0.153) &
[1] \cellcolor{blue!20}0.004 (0.009) \\
&\textsc{\textbf{DeepArch}}&
[1] \cellcolor{blue!20}0.000 (0.000) &
[3] 0.035 (0.089) &
[4] 0.050 (0.114) &
[4] 0.111 (0.207) &
[1] \cellcolor{blue!20}0.000 (0.002) &
[4] 0.047 (0.095) &
[3] 0.014 (0.075) &
[1] \cellcolor{blue!20}0.000 (0.002) &
[4] 0.157 (0.253) &
[3] 0.015 (0.075) &
[2] \cellcolor{teal!20}0.002 (0.011) &
[1] \cellcolor{blue!20}0.000 (0.000) \\
&\textsc{\textbf{PostgreSQL}}&
[2] \cellcolor{teal!20}0.203 (0.172) &
[3] 0.230 (0.151) &
[3] 0.237 (0.157) &
[2] \cellcolor{teal!20}0.175 (0.158) &
[5] 1.000 (0.000) &
[3] 0.271 (0.260) &
[1] \cellcolor{blue!20}0.116 (0.071) &
[3] 0.230 (0.148) &
[4] 0.442 (0.242) &
[3] 0.225 (0.163) &
[2] \cellcolor{teal!20}0.165 (0.120) &
[1] \cellcolor{blue!20}0.123 (0.108) \\
&\textsc{\textbf{JavaGC}}&
[2] \cellcolor{teal!20}0.138 (0.126) &
[3] 0.149 (0.127) &
[3] 0.181 (0.150) &
[4] 0.266 (0.190) &
[5] 1.000 (0.000) &
[3] 0.210 (0.167) &
[6] \textcolor{red}{\ding{55}} &
[6] \textcolor{red}{\ding{55}} &
[6] \textcolor{red}{\ding{55}} &
[3] 0.197 (0.167) &
[1] \cellcolor{blue!20}0.090 (0.074) &
[2] \cellcolor{teal!20}0.130 (0.000) \\
&\textsc{\textbf{Storm}}&
[2] \cellcolor{teal!20}0.007 (0.009) &
[8] 0.195 (0.119) &
[8] 0.194 (0.122) &
[7] 0.191 (0.200) &
[7] 0.154 (0.122) &
[9] 0.292 (0.204) &
[4] 0.024 (0.047) &
[1] \cellcolor{blue!20}0.003 (0.007) &
[5] 0.028 (0.069) &
[6] 0.053 (0.181) &
[3] 0.014 (0.038) &
[3] 0.016 (0.057) \\
&\textsc{\textbf{x264}}&
[1] \cellcolor{blue!20}0.247 (0.073) &
[3] 0.285 (0.063) &
[3] 0.288 (0.078) &
[4] 0.339 (0.131) &
[5] 1.000 (0.000) &
[2] \cellcolor{teal!20}0.271 (0.055) &
[6] \textcolor{red}{\ding{55}} &
[6] \textcolor{red}{\ding{55}} &
[6] \textcolor{red}{\ding{55}} &
[1] \cellcolor{blue!20}0.250 (0.127) &
[1] \cellcolor{blue!20}0.252 (0.092) &
[3] 0.280 (0.000) \\
&\textsc{\textbf{Redis}}&
[1] \cellcolor{blue!20}0.389 (0.151) &
[3] 0.492 (0.189) &
[3] 0.492 (0.189) &
[4] 0.608 (0.164) &
[2] \cellcolor{teal!20}0.437 (0.128) &
[5] 0.701 (0.142) &
[7] \textcolor{red}{\ding{55}} &
[7] \textcolor{red}{\ding{55}} &
[7] \textcolor{red}{\ding{55}} &
[2] \cellcolor{teal!20}0.428 (0.178) &
[4] 0.624 (0.176) &
[6] 0.795 (0.121) \\
&\textsc{\textbf{HSQLDB}}&
[4] 0.020 (0.028) &
[4] 0.027 (0.022) &
[5] 0.028 (0.023) &
[5] 0.043 (0.037) &
[8] 1.000 (0.000) &
[4] 0.024 (0.020) &
[1] \cellcolor{blue!20}0.005 (0.014) &
[4] 0.018 (0.023) &
[7] 0.083 (0.241) &
[6] 0.049 (0.174) &
[3] 0.010 (0.016) &
[2] \cellcolor{teal!20}0.005 (0.011) \\
&\textsc{\textbf{LLVM}}&
[2] \cellcolor{teal!20}0.012 (0.018) &
[5] 0.155 (0.117) &
[5] 0.152 (0.115) &
[4] 0.050 (0.055) &
[4] 0.044 (0.029) &
[6] 0.280 (0.259) &
[1] \cellcolor{blue!20}0.009 (0.005) &
[2] \cellcolor{teal!20}0.013 (0.019) &
[4] 0.081 (0.133) &
[2] \cellcolor{teal!20}0.016 (0.026) &
[1] \cellcolor{blue!20}0.008 (0.012) &
[3] 0.024 (0.026) \\

    \hline
\multicolumn{2}{c}{\textbf{All systems at $B=50$}}&\textbf{[1.8] 0.109 (0.083)}&\textbf{[3.9] 0.176 (0.109)}&\textbf{[4.1] 0.181 (0.116)}&\textbf{[4.6] 0.234 (0.176)}&\textbf{[4.8] 0.570 (0.045)}&\textbf{[4.3] 0.243 (0.167)}&\textbf{[3.5] 0.319 (0.068)}&\textbf{[3.4] 0.307 (0.066)}&\textbf{[4.6] 0.367 (0.135)}&\textbf{[3.3] 0.150 (0.147)}&\textbf{[2.7] 0.146 (0.108)}&\textbf{[2.5] 0.208 (0.043)}\\

    \hline
\multirow{14}{*}{$B=100$}

&\textsc{\textbf{7z}}&
[1] \cellcolor{blue!20}0.012 (0.007) &
[2] \cellcolor{teal!20}0.074 (0.116) &
[3] 0.075 (0.117) &
[5] 0.242 (0.279) &
[6] 0.691 (0.002) &
[4] 0.204 (0.297) &
[4] 0.191 (0.265) &
[2] \cellcolor{teal!20}0.040 (0.116) &
[4] 0.148 (0.265) &
[5] 0.245 (0.308) &
[3] 0.118 (0.203) &
[7] 0.990 (0.054) \\
&\textsc{\textbf{DConvert}}&
[2] \cellcolor{teal!20}0.032 (0.052) &
[5] 0.151 (0.103) &
[5] 0.153 (0.091) &
[6] 0.301 (0.285) &
[7] 0.694 (0.011) &
[5] 0.193 (0.232) &
[1] \cellcolor{blue!20}0.023 (0.051) &
[2] \cellcolor{teal!20}0.050 (0.061) &
[3] 0.070 (0.071) &
[4] 0.112 (0.238) &
[4] 0.102 (0.168) &
[1] \cellcolor{blue!20}0.000 (0.000) \\
&\textsc{\textbf{ExaStencils}}&
[1] \cellcolor{blue!20}0.062 (0.076) &
[3] 0.109 (0.054) &
[3] 0.105 (0.049) &
[3] 0.123 (0.093) &
[4] 0.755 (0.135) &
[3] 0.101 (0.068) &
[1] \cellcolor{blue!20}0.067 (0.081) &
[1] \cellcolor{blue!20}0.056 (0.073) &
[3] 0.114 (0.078) &
[3] 0.102 (0.085) &
[2] \cellcolor{teal!20}0.074 (0.075) &
[1] \cellcolor{blue!20}0.038 (0.058) \\
&\textsc{\textbf{BDB-C}}&
[1] \cellcolor{blue!20}0.007 (0.013) &
[5] 0.027 (0.020) &
[4] 0.027 (0.019) &
[7] 0.149 (0.194) &
[2] \cellcolor{teal!20}0.010 (0.018) &
[5] 0.067 (0.097) &
[8] 0.241 (0.168) &
[5] 0.029 (0.066) &
[7] 0.106 (0.215) &
[3] 0.012 (0.022) &
[6] 0.086 (0.122) &
[1] \cellcolor{blue!20}0.000 (0.000) \\
&\textsc{\textbf{DeepArch}}&
[1] \cellcolor{blue!20}0.000 (0.000) &
[3] 0.003 (0.006) &
[2] \cellcolor{teal!20}0.003 (0.006) &
[4] 0.099 (0.208) &
[1] \cellcolor{blue!20}0.000 (0.000) &
[4] 0.047 (0.095) &
[2] \cellcolor{teal!20}0.000 (0.002) &
[1] \cellcolor{blue!20}0.000 (0.000) &
[5] 0.157 (0.253) &
[4] 0.014 (0.075) &
[1] \cellcolor{blue!20}0.000 (0.000) &
[1] \cellcolor{blue!20}0.000 (0.000) \\
&\textsc{\textbf{PostgreSQL}}&
[1] \cellcolor{blue!20}0.064 (0.086) &
[3] 0.139 (0.102) &
[3] 0.142 (0.098) &
[3] 0.165 (0.158) &
[6] 1.000 (0.000) &
[4] 0.245 (0.257) &
[1] \cellcolor{blue!20}0.049 (0.040) &
[2] \cellcolor{teal!20}0.100 (0.124) &
[5] 0.442 (0.242) &
[3] 0.152 (0.141) &
[1] \cellcolor{blue!20}0.093 (0.095) &
[1] \cellcolor{blue!20}0.060 (0.059) \\
&\textsc{\textbf{JavaGC}}&
[1] \cellcolor{blue!20}0.056 (0.038) &
[2] \cellcolor{teal!20}0.101 (0.095) &
[2] \cellcolor{teal!20}0.099 (0.093) &
[3] 0.241 (0.176) &
[4] 1.000 (0.000) &
[2] \cellcolor{teal!20}0.101 (0.111) &
[5] \textcolor{red}{\ding{55}} &
[5] \textcolor{red}{\ding{55}} &
[5] \textcolor{red}{\ding{55}} &
[2] \cellcolor{teal!20}0.127 (0.131) &
[1] \cellcolor{blue!20}0.052 (0.040) &
[3] 0.130 (0.000) \\
&\textsc{\textbf{Storm}}&
[1] \cellcolor{blue!20}0.000 (0.000) &
[6] 0.149 (0.110) &
[6] 0.141 (0.106) &
[6] 0.151 (0.139) &
[4] 0.043 (0.071) &
[6] 0.143 (0.121) &
[3] 0.008 (0.010) &
[1] \cellcolor{blue!20}0.000 (0.000) &
[3] 0.028 (0.069) &
[5] 0.047 (0.182) &
[2] \cellcolor{teal!20}0.001 (0.004) &
[2] \cellcolor{teal!20}0.001 (0.004) \\
&\textsc{\textbf{x264}}&
[1] \cellcolor{blue!20}0.220 (0.062) &
[2] \cellcolor{teal!20}0.236 (0.059) &
[1] \cellcolor{blue!20}0.224 (0.064) &
[4] 0.328 (0.133) &
[5] 1.000 (0.000) &
[2] \cellcolor{teal!20}0.252 (0.053) &
[6] \textcolor{red}{\ding{55}} &
[6] \textcolor{red}{\ding{55}} &
[6] \textcolor{red}{\ding{55}} &
[2] \cellcolor{teal!20}0.238 (0.106) &
[1] \cellcolor{blue!20}0.212 (0.086) &
[3] 0.280 (0.000) \\
&\textsc{\textbf{Redis}}&
[1] \cellcolor{blue!20}0.307 (0.135) &
[2] \cellcolor{teal!20}0.362 (0.142) &
[2] \cellcolor{teal!20}0.362 (0.142) &
[3] 0.588 (0.151) &
[2] \cellcolor{teal!20}0.357 (0.120) &
[3] 0.590 (0.102) &
[5] \textcolor{red}{\ding{55}} &
[5] \textcolor{red}{\ding{55}} &
[5] \textcolor{red}{\ding{55}} &
[2] \cellcolor{teal!20}0.389 (0.183) &
[3] 0.599 (0.149) &
[4] 0.795 (0.121) \\
&\textsc{\textbf{HSQLDB}}&
[3] 0.008 (0.015) &
[4] 0.014 (0.017) &
[4] 0.014 (0.017) &
[4] 0.041 (0.038) &
[6] 1.000 (0.000) &
[4] 0.013 (0.015) &
[1] \cellcolor{blue!20}0.000 (0.000) &
[2] \cellcolor{teal!20}0.006 (0.012) &
[5] 0.082 (0.242) &
[4] 0.043 (0.175) &
[2] \cellcolor{teal!20}0.007 (0.012) &
[1] \cellcolor{blue!20}0.000 (0.000) \\
&\textsc{\textbf{LLVM}}&
[1] \cellcolor{blue!20}0.000 (0.000) &
[8] 0.075 (0.056) &
[7] 0.074 (0.049) &
[7] 0.049 (0.055) &
[4] 0.005 (0.009) &
[9] 0.280 (0.259) &
[3] 0.001 (0.002) &
[3] 0.000 (0.002) &
[8] 0.081 (0.133) &
[6] 0.009 (0.025) &
[2] \cellcolor{teal!20}0.000 (0.002) &
[5] 0.006 (0.005) \\

  \hline
\multicolumn{2}{c}{\textbf{All systems at $B=100$}}&\textbf{[1.2] 0.064 (0.040)}&\textbf{[3.8] 0.120 (0.073)}&\textbf{[3.5] 0.118 (0.071)}&\textbf{[4.6] 0.206 (0.159)}&\textbf{[4.2] 0.546 (0.030)}&\textbf{[4.2] 0.186 (0.142)}&\textbf{[3.3] 0.298 (0.052)}&\textbf{[2.9] 0.273 (0.038)}&\textbf{[4.9] 0.352 (0.131)}&\textbf{[3.6] 0.124 (0.139)}&\textbf{[2.3] 0.112 (0.080)}&\textbf{[2.5] 0.192 (0.025)}\\
\hline
\multirow{14}{*}{$B=150$}

&\textsc{\textbf{7z}}&
[1] \cellcolor{blue!20}0.009 (0.007) &
[3] 0.041 (0.020) &
[2] \cellcolor{teal!20}0.039 (0.019) &
[7] 0.242 (0.279) &
[8] 0.691 (0.002) &
[5] 0.179 (0.285) &
[5] 0.142 (0.241) &
[2] \cellcolor{teal!20}0.016 (0.014) &
[4] 0.080 (0.200) &
[6] 0.221 (0.300) &
[3] 0.063 (0.125) &
[9] 0.990 (0.054) \\
&\textsc{\textbf{DConvert}}&
[1] \cellcolor{blue!20}0.013 (0.032) &
[5] 0.110 (0.096) &
[4] 0.110 (0.087) &
[6] 0.300 (0.285) &
[7] 0.690 (0.000) &
[5] 0.129 (0.174) &
[2] \cellcolor{teal!20}0.015 (0.045) &
[2] \cellcolor{teal!20}0.040 (0.055) &
[3] 0.059 (0.061) &
[5] 0.112 (0.238) &
[4] 0.097 (0.169) &
[1] \cellcolor{blue!20}0.000 (0.000) \\
&\textsc{\textbf{ExaStencils}}&
[2] \cellcolor{teal!20}0.058 (0.077) &
[4] 0.083 (0.040) &
[4] 0.085 (0.035) &
[6] 0.123 (0.093) &
[7] 0.755 (0.135) &
[5] 0.087 (0.071) &
[2] \cellcolor{teal!20}0.051 (0.073) &
[2] \cellcolor{teal!20}0.053 (0.071) &
[6] 0.114 (0.078) &
[5] 0.096 (0.088) &
[3] 0.073 (0.075) &
[1] \cellcolor{blue!20}0.025 (0.057) \\
&\textsc{\textbf{BDB-C}}&
[2] \cellcolor{teal!20}0.001 (0.006) &
[5] 0.014 (0.015) &
[5] 0.017 (0.018) &
[8] 0.149 (0.194) &
[2] \cellcolor{teal!20}0.005 (0.012) &
[5] 0.054 (0.074) &
[9] 0.207 (0.175) &
[3] 0.007 (0.013) &
[7] 0.085 (0.208) &
[4] 0.008 (0.013) &
[6] 0.074 (0.110) &
[1] \cellcolor{blue!20}0.000 (0.000) \\
&\textsc{\textbf{DeepArch}}&
[1] \cellcolor{blue!20}0.000 (0.000) &
[3] 0.002 (0.006) &
[3] 0.002 (0.006) &
[4] 0.099 (0.208) &
[1] \cellcolor{blue!20}0.000 (0.000) &
[4] 0.047 (0.095) &
[2] \cellcolor{teal!20}0.000 (0.002) &
[1] \cellcolor{blue!20}0.000 (0.000) &
[5] 0.157 (0.253) &
[4] 0.014 (0.075) &
[1] \cellcolor{blue!20}0.000 (0.000) &
[1] \cellcolor{blue!20}0.000 (0.000) \\
&\textsc{\textbf{PostgreSQL}}&
[1] \cellcolor{blue!20}0.042 (0.067) &
[4] 0.114 (0.088) &
[4] 0.114 (0.085) &
[4] 0.165 (0.158) &
[7] 1.000 (0.000) &
[5] 0.245 (0.257) &
[1] \cellcolor{blue!20}0.034 (0.037) &
[2] \cellcolor{teal!20}0.062 (0.092) &
[6] 0.442 (0.242) &
[4] 0.139 (0.142) &
[3] 0.076 (0.091) &
[1] \cellcolor{blue!20}0.030 (0.037) \\
&\textsc{\textbf{JavaGC}}&
[1] \cellcolor{blue!20}0.049 (0.031) &
[2] \cellcolor{teal!20}0.081 (0.080) &
[2] \cellcolor{teal!20}0.080 (0.078) &
[3] 0.241 (0.176) &
[4] 1.000 (0.000) &
[1] \cellcolor{blue!20}0.072 (0.074) &
[5] \textcolor{red}{\ding{55}} &
[5] \textcolor{red}{\ding{55}} &
[5] \textcolor{red}{\ding{55}} &
[2] \cellcolor{teal!20}0.086 (0.102) &
[1] \cellcolor{blue!20}0.050 (0.040) &
[3] 0.130 (0.000) \\
&\textsc{\textbf{Storm}}&
[1] \cellcolor{blue!20}0.000 (0.000) &
[4] 0.089 (0.076) &
[4] 0.098 (0.093) &
[6] 0.150 (0.140) &
[2] 0.020 (0.029) &
[5] 0.122 (0.116) &
[2] \cellcolor{teal!20}0.007 (0.009) &
[1] \cellcolor{blue!20}0.000 (0.000) &
[3] 0.028 (0.069) &
[3] 0.047 (0.182) &
[1] \cellcolor{blue!20}0.000 (0.000) &
[1] \cellcolor{blue!20}0.000 (0.000) \\
&\textsc{\textbf{x264}}&
[2] \cellcolor{teal!20}0.204 (0.060) &
[1] \cellcolor{blue!20}0.198 (0.069) &
[2] \cellcolor{teal!20}0.202 (0.075) &
[5] 0.328 (0.133) &
[6] 1.000 (0.000) &
[3] 0.225 (0.060) &
[7] \textcolor{red}{\ding{55}} &
[7] \textcolor{red}{\ding{55}} &
[7] \textcolor{red}{\ding{55}} &
[3] 0.235 (0.104) &
[2] \cellcolor{teal!20}0.203 (0.081) &
[4] 0.280 (0.000) \\
&\textsc{\textbf{Redis}}&
[1] \cellcolor{blue!20}0.263 (0.142) &
[2] \cellcolor{teal!20}0.311 (0.133) &
[2] \cellcolor{teal!20}0.311 (0.133) &
[5] 0.588 (0.151) &
[1] \cellcolor{blue!20}0.291 (0.119) &
[4] 0.537 (0.108) &
[7] \textcolor{red}{\ding{55}} &
[7] \textcolor{red}{\ding{55}} &
[7] \textcolor{red}{\ding{55}} &
[3] 0.385 (0.181) &
[5] 0.590 (0.145) &
[6] 0.795 (0.121) \\
&\textsc{\textbf{HSQLDB}}&
[2] \cellcolor{teal!20}0.003 (0.008) &
[5] 0.008 (0.012) &
[4] 0.008 (0.012) &
[6] 0.041 (0.038) &
[8] 1.000 (0.000) &
[6] 0.012 (0.014) &
[1] \cellcolor{blue!20}0.000 (0.000) &
[4] 0.006 (0.012) &
[7] 0.082 (0.242) &
[6] 0.043 (0.175) &
[3] 0.005 (0.009) &
[1] \cellcolor{blue!20}0.000 (0.000) \\
&\textsc{\textbf{LLVM}}&
[1] \cellcolor{blue!20}0.000 (0.000) &
[5] 0.052 (0.041) &
[5] 0.051 (0.040) &
[5] 0.049 (0.055) &
[2] \cellcolor{teal!20}0.001 (0.002) &
[6] 0.280 (0.259) &
[1] \cellcolor{blue!20}0.000 (0.002) &
[1] \cellcolor{blue!20}0.000 (0.000) &
[5] 0.081 (0.133) &
[4] 0.009 (0.025) &
[1] \cellcolor{blue!20}0.000 (0.000) &
[3] 0.002 (0.004) \\

  \hline
\multicolumn{2}{c}{\textbf{All systems at $B=150$}}&\textbf{[1.3] 0.053 (0.036)}&\textbf{[3.6] 0.092 (0.056)}&\textbf{[3.4] 0.093 (0.057)}&\textbf{[5.4] 0.206 (0.159)}&\textbf{[4.6] 0.538 (0.025)}&\textbf{[4.5] 0.166 (0.132)}&\textbf{[3.7] 0.288 (0.049)}&\textbf{[3.1] 0.265 (0.021)}&\textbf{[5.4] 0.344 (0.124)}&\textbf{[4.1] 0.116 (0.135)}&\textbf{[2.8] 0.103 (0.070)}&\textbf{[2.7] 0.188 (0.023)}\\
    \hline
\multirow{14}{*}{$B=200$}

&\textsc{\textbf{7z}}&
[1] \cellcolor{blue!20}0.006 (0.005) &
[4] 0.036 (0.021) &
[3] 0.035 (0.020) &
[7] 0.242 (0.279) &
[8] 0.691 (0.002) &
[5] 0.121 (0.217) &
[5] 0.111 (0.223) &
[3] 0.012 (0.006) &
[2] \cellcolor{teal!20}0.012 (0.008) &
[6] 0.218 (0.296) &
[5] 0.060 (0.125) &
[9] 0.990 (0.054) \\
&\textsc{\textbf{DConvert}}&
[2] \cellcolor{teal!20}0.013 (0.032) &
[4] 0.076 (0.072) &
[5] 0.086 (0.077) &
[7] 0.300 (0.285) &
[8] 0.690 (0.000) &
[6] 0.121 (0.174) &
[1] \cellcolor{blue!20}0.000 (0.000) &
[2] \cellcolor{teal!20}0.036 (0.051) &
[3] 0.059 (0.061) &
[6] 0.112 (0.238) &
[5] 0.091 (0.168) &
[1] \cellcolor{blue!20}0.000 (0.000) \\
&\textsc{\textbf{ExaStencils}}&
[2] \cellcolor{teal!20}0.057 (0.077) &
[3] 0.069 (0.026) &
[4] 0.073 (0.027) &
[5] 0.123 (0.093) &
[6] 0.755 (0.135) &
[4] 0.080 (0.072) &
[2] \cellcolor{teal!20}0.048 (0.072) &
[2] \cellcolor{teal!20}0.048 (0.070) &
[5] 0.114 (0.078) &
[4] 0.095 (0.088) &
[4] 0.070 (0.075) &
[1] \cellcolor{blue!20}0.023 (0.058) \\
&\textsc{\textbf{BDB-C}}&
[1] \cellcolor{blue!20}0.000 (0.000) &
[5] 0.012 (0.015) &
[6] 0.012 (0.015) &
[8] 0.149 (0.194) &
[2] \cellcolor{teal!20}0.002 (0.006) &
[6] 0.039 (0.046) &
[8] 0.184 (0.177) &
[3] 0.005 (0.011) &
[8] 0.083 (0.208) &
[4] 0.006 (0.012) &
[7] 0.071 (0.109) &
[1] \cellcolor{blue!20}0.000 (0.000) \\
&\textsc{\textbf{DeepArch}}&
[1] \cellcolor{blue!20}0.000 (0.000) &
[2] \cellcolor{teal!20}0.000 (0.002) &
[2] \cellcolor{teal!20}0.000 (0.002) &
[3] 0.099 (0.208) &
[1] \cellcolor{blue!20}0.000 (0.000) &
[3] 0.047 (0.095) &
[2] \cellcolor{teal!20}0.000 (0.002) &
[1] \cellcolor{blue!20}0.000 (0.000) &
[4] 0.157 (0.253) &
[3] 0.014 (0.075) &
[1] \cellcolor{blue!20}0.000 (0.000) &
[1] \cellcolor{blue!20}0.000 (0.000) \\
&\textsc{\textbf{PostgreSQL}}&
[2] \cellcolor{teal!20}0.027 (0.051) &
[4] 0.103 (0.084) &
[4] 0.090 (0.075) &
[5] 0.165 (0.158) &
[8] 1.000 (0.000) &
[6] 0.245 (0.257) &
[1] \cellcolor{blue!20}0.020 (0.034) &
[2] \cellcolor{teal!20}0.050 (0.081) &
[7] 0.442 (0.242) &
[4] 0.139 (0.142) &
[3] 0.067 (0.087) &
[1] \cellcolor{blue!20}0.025 (0.031) \\
&\textsc{\textbf{JavaGC}}&
[1] \cellcolor{blue!20}0.042 (0.018) &
[4] 0.056 (0.034) &
[3] 0.052 (0.023) &
[6] 0.241 (0.176) &
[7] 1.000 (0.000) &
[5] 0.058 (0.063) &
[8] \textcolor{red}{\ding{55}} &
[8] \textcolor{red}{\ding{55}} &
[8] \textcolor{red}{\ding{55}} &
[5] 0.063 (0.086) &
[2] \cellcolor{teal!20}0.046 (0.040) &
[6] 0.130 (0.000) \\
&\textsc{\textbf{Storm}}&
[1] \cellcolor{blue!20}0.000 (0.000) &
[5] 0.064 (0.072) &
[5] 0.055 (0.065) &
[6] 0.150 (0.140) &
[3] 0.007 (0.009) &
[6] 0.113 (0.115) &
[2] \cellcolor{teal!20}0.003 (0.007) &
[1] \cellcolor{blue!20}0.000 (0.000) &
[3] 0.028 (0.069) &
[4] 0.047 (0.182) &
[1] \cellcolor{blue!20}0.000 (0.000) &
[1] \cellcolor{blue!20}0.000 (0.000) \\
&\textsc{\textbf{x264}}&
[2] \cellcolor{teal!20}0.197 (0.051) &
[1] \cellcolor{blue!20}0.188 (0.070) &
[1] \cellcolor{blue!20}0.184 (0.075) &
[5] 0.328 (0.133) &
[6] 1.000 (0.000) &
[3] 0.202 (0.065) &
[7] \textcolor{red}{\ding{55}} &
[7] \textcolor{red}{\ding{55}} &
[7] \textcolor{red}{\ding{55}} &
[3] 0.235 (0.104) &
[3] 0.203 (0.081) &
[4] 0.280 (0.000) \\
&\textsc{\textbf{Redis}}&
[1] \cellcolor{blue!20}0.236 (0.138) &
[2] \cellcolor{teal!20}0.285 (0.111) &
[2] \cellcolor{teal!20}0.285 (0.111) &
[5] 0.588 (0.151) &
[1] \cellcolor{blue!20}0.232 (0.109) &
[4] 0.492 (0.117) &
[7] \textcolor{red}{\ding{55}} &
[7] \textcolor{red}{\ding{55}} &
[7] \textcolor{red}{\ding{55}} &
[3] 0.385 (0.181) &
[5] 0.590 (0.145) &
[6] 0.795 (0.121) \\
&\textsc{\textbf{HSQLDB}}&
[2] \cellcolor{teal!20}0.003 (0.008) &
[4] 0.007 (0.011) &
[5] 0.007 (0.011) &
[6] 0.041 (0.038) &
[8] 1.000 (0.000) &
[6] 0.010 (0.013) &
[1] \cellcolor{blue!20}0.000 (0.000) &
[4] 0.005 (0.011) &
[7] 0.082 (0.242) &
[6] 0.043 (0.175) &
[3] 0.004 (0.008) &
[1] \cellcolor{blue!20}0.000 (0.000) \\
&\textsc{\textbf{LLVM}}&
[1] \cellcolor{blue!20}0.000 (0.000) &
[4] 0.046 (0.037) &
[4] 0.047 (0.039) &
[4] 0.049 (0.055) &
[1] \cellcolor{blue!20}0.000 (0.000) &
[5] 0.280 (0.259) &
[2] \cellcolor{teal!20}0.000 (0.002) &
[1] \cellcolor{blue!20}0.000 (0.000) &
[4] 0.081 (0.133) &
[3] 0.009 (0.025) &
[1] \cellcolor{blue!20}0.000 (0.000) &
[3] 0.001 (0.002) \\

\hline
\multicolumn{2}{c}{\textbf{All systems at $B=200$}}&\textbf{[1.4] 0.048 (0.032)}&\textbf{[3.5] 0.079 (0.046)}&\textbf{[3.7] 0.077 (0.045)}&\textbf{[5.6] 0.206 (0.159)}&\textbf{[4.9] 0.531 (0.022)}&\textbf{[4.9] 0.151 (0.124)}&\textbf{[3.8] 0.280 (0.043)}&\textbf{[3.4] 0.263 (0.019)}&\textbf{[5.4] 0.338 (0.108)}&\textbf{[4.2] 0.114 (0.134)}&\textbf{[3.3] 0.100 (0.070)}&\textbf{[2.9] 0.187 (0.022)}\\

\midrule \midrule
\multicolumn{2}{c}{\textbf{All systems/budgets}}&\textbf{[1.5] 0.069 (0.048)}&\textbf{[3.7] 0.117 (0.071)}&\textbf{[3.7] 0.117 (0.072)}&\textbf{[5.0] 0.213 (0.163)}&\textbf{[4.6] 0.546 (0.030)}&\textbf{[4.5] 0.186 (0.141)}&\textbf{[3.6] 0.297 (0.053)}&\textbf{[3.2] 0.277 (0.036)}&\textbf{[5.1] 0.350 (0.124)}&\textbf{[3.8] 0.126 (0.139)}&\textbf{[2.8] 0.115 (0.082)}&\textbf{[2.6] 0.194 (0.028)}\\

    \bottomrule
  \end{tabular}
\end{adjustbox}
\vspace{-0.2cm}
\end{table*}

For comparing multiple tuners, we leverage the Scott-Knott ESD test \cite{DBLP:journals/tse/Herbold17}. In a nutshell, it first ranks the approaches based on the mean performance scores and then iteratively partitions this ordered list into statistically distinct subgroups, which are determined by maximizing the inter-group mean square difference \(\Delta\) and their effect sizes. For example, for three approaches $A$, $B$, and $C$, the Scott-Knott ESD test may yield two groups: $\{A, B\}$ with rank 1 and $\{C\}$ with rank 2, meaning that $A$ and $B$ are statistically 
similar but they are both significantly better than $C$.  Compared with other methods such as the Kruskal-Wallis test \cite{mckight2010kruskal}, Scott-Knott ESD overcomes the confounding factor of overlapping groups~\cite{DBLP:journals/tse/Tantithamthavorn19,DBLP:conf/icse/GhotraMH15,DBLP:journals/tse/MittasA13} while it does not require post-hoc correction and can indicate better approaches.

\section{Results}
\label{sec:results}

\subsection{RQ1: Effectiveness}



\subsubsection{Method}

For \textbf{RQ1}, we compare all 10 state-of-the-art and baseline tuners mentioned in Section~\ref{sec:sota} under $12$ systems with four budgets, leading to $12\times4=48$ cases. For each case, we use Scott-Knott ESD to rank the tuners over 30 runs and highlight the one(s) with the best rank, meaning that they are statistically better than the others. To ensure consistency and ease of exposition, the performance is normalized across the systems for each budget. 

\subsubsection{Results}

As from Table~\ref{tab: performance}, we see that \tuner~perform remarkably better and more stable than the others, achieving an overall rank of 1.5, within which it is ranked the best or second best for 93\% (45/48) cases (the best for 30/48 cases). This significantly outperforms the overall second best tuner, i.e., \texttt{HEBO}, which has an overall rank of 2.6 and it is ranked the best or second best for 58\% (28/48) cases only (25/48 cases as the best). \texttt{HEBO} is also unstable as it easily leads to devastating results: for the remaining 20 cases, it is commonly ranked as one of the worst. The improvements of \tuner~is overall significant, i.e., up to a few orders of magnitude better than the second best tuner. There are cases where the tuners cannot complete one run even after $24$ hours due to their greedy search assumption, which fails to consider the systems with a large configuration space. For example, \texttt{FLASH} needs to traverse the entire search space at each iteration, which makes it struggle for complex systems like \textsc{JavaGC}. In summary, we say


\begin{quotebox}
   \noindent
   \textit{\textbf{RQ1:} \tuner~performs considerably better and more stable than the state-of-the-art tuners, achieving an overall rank of 1.5---42\% better than the second best tuner---with significant performance improvement.}
\end{quotebox}



\subsection{RQ2: Ablation Study}



\begin{figure*}[!t]
\centering

\begin{subfigure}{0.15\textwidth}
  \centering
    \begin{tikzpicture}
    \begin{axis}[
        axis x line  = bottom,
        axis y line  = left,
        width=4cm, height=4cm,
        xlabel={Budget},
        ylabel={Normalized performance},
        grid=minor,
        ymax=0.09,
        ymin=0,
        legend columns=2, 
        legend style={draw=none,fill=none,at={(1.0,1)}, anchor=south east, legend cell align=left}
    ]

    \addplot[
        blue,
        mark=none,
          thick
    ] coordinates {
(50,0.07106666666666667)
(100,0.013400000000000002)
(150,0.010133333333333334)
(200,0.007633333333333335)
    };

    \addplot[
        red,
        dashed,
        mark=none,
          thick
    ] coordinates {
(50,0.08403333333333335)
(100,0.03626666666666667)
(150,0.029033333333333335)
(200,0.017433333333333335)
    };

    \addplot[
        fill=blue!30,
        draw=none,
        fill opacity=0.3
    ] coordinates {
(50,0.23545197284969782)
(100,0.019001190349678662)
(150,0.016070410803679235)
(200,0.012636554517952522)
(200,0.002630112148714148)
(150,0.004196255862987433)
(100,0.007798809650321343)
(50,-0.09331863951636449)
    } \closedcycle;

    \addplot[
        fill=red!30,
        draw=none,
        fill opacity=0.3
    ] coordinates {
(50,0.24536869489083224)
(100,0.07289173397794345)
(150,0.062383479093467906)
(200,0.028713650507755144)
(200,0.006153016158911529)
(150,-0.0043168124268012364)
(100,-0.0003584006446101101)
(50,-0.07730202822416554)
    } \closedcycle;

    \end{axis}
\end{tikzpicture}
     \vspace{-0.6cm}
   \caption{\textsc{7z}}
\end{subfigure}
~\hfill
\begin{subfigure}{0.15\textwidth}
  \centering
    \begin{tikzpicture}
    \begin{axis}[
        axis x line  = bottom,
        axis y line  = left,
        width=4cm, height=4cm,
        xlabel={Budget},
        ylabel={Normalized performance},
        grid=minor,
        scaled y ticks={base 10:2},
         ymax=0.145,
        ymin=-0.01,
        legend columns=2, 
        legend style={draw=none,fill=none,at={(1.0,1)}, anchor=south east, legend cell align=left}
    ]

    \addplot[
        blue,
        mark=none,
         thick
    ] coordinates {
(50,0.07783333333333334)
(100,0.03253333333333333)
(150,0.012866666666666665)
(200,0.012866666666666665)
    };

    \addplot[
        red,
        dashed,
        mark=none,
         thick
    ] coordinates {
(50,0.14179999999999998)
(100,0.05473333333333333)
(150,0.04536666666666667)
(200,0.04366666666666666)
    };

    \addplot[
        fill=blue!30,
        draw=none,
        fill opacity=0.3
    ] coordinates {
(50,0.15045109210409746)
(100,0.0846791118990257)
(150,0.045926948308271334)
(200,0.045926948308271334)
(200,-0.020193614974938004)
(150,-0.020193614974938004)
(100,-0.01961244523235904)
(50,0.005215574562569214)
    } \closedcycle;

    \addplot[
        fill=red!30,
        draw=none,
        fill opacity=0.3
    ] coordinates {
(50,0.2976341853809148)
(100,0.11107688270285349)
(150,0.09649492586731703)
(200,0.09416837180839649)
(200,-0.006835038475063178)
(150,-0.0057615925339836935)
(100,-0.001610216036186822)
(50,-0.01403418538091486)
    } \closedcycle;

    \end{axis}
\end{tikzpicture}
    \vspace{-0.6cm}
   \caption{\textsc{DConvert}}
\end{subfigure}
~\hfill
\begin{subfigure}{0.15\textwidth}
  \centering
    \begin{tikzpicture}
    \begin{axis}[
        axis x line  = bottom,
        axis y line  = left,
        width=4cm, height=4cm,
        xlabel={Budget},
        ylabel={Normalized performance},
        grid=minor,
        legend columns=2, 
         ymax=0.095,
        ymin=0.05,
        legend style={draw=none,fill=none,at={(1.0,1)}, anchor=south east, legend cell align=left}
    ]

    \addplot[
        blue,
        mark=none,
         thick
    ] coordinates {
(50,0.08796666666666665)
(100,0.06319999999999999)
(150,0.05883333333333333)
(200,0.0579)
    };

    \addplot[
        red,
        dashed,
        mark=none,
         thick
    ] coordinates {
(50,0.09416666666666665)
(100,0.07266666666666664)
(150,0.061766666666666664)
(200,0.053599999999999995)
    };

    \addplot[
        fill=blue!30,
        draw=none,
        fill opacity=0.3
    ] coordinates {
(50,0.16648948005374686)
(100,0.13909835307831125)
(150,0.13581302040870552)
(200,0.13493261214143873)
(200,-0.019132612141438722)
(150,-0.018146353742038858)
(100,-0.01269835307831127)
(50,0.00944385327958644)
    } \closedcycle;

    \addplot[
        fill=red!30,
        draw=none,
        fill opacity=0.3
    ] coordinates {
(50,0.17008420233677535)
(100,0.14844191914984145)
(150,0.13721452209602192)
(200,0.12558823978030487)
(200,-0.018388239780304863)
(150,-0.013681188762688576)
(100,-0.0031085858165081637)
(50,0.018249130996557963)
    } \closedcycle;

    \end{axis}
\end{tikzpicture}
      \vspace{-0.6cm}
   \caption{\textsc{ExaStencils}}
\end{subfigure}
~\hfill
\begin{subfigure}{0.15\textwidth}
  \centering
    \begin{tikzpicture}
    \begin{axis}[
        axis x line  = bottom,
        axis y line  = left,
        width=4cm, height=4cm,
        xlabel={Budget},
        ylabel={Normalized performance},
        grid=minor,
        ymax=0.12,
        scaled y ticks={base 10:2},
         ymin=-0.005,
        legend columns=2, 
        legend style={draw=none,fill=none,at={(1.0,1)}, anchor=south east, legend cell align=left}
    ]

    \addplot[
        blue,
        mark=none,
         thick
    ] coordinates {
(50,0.0542)
(100,0.0076)
(150,0.0013333333333333335)
(200,0.0001)
    };

    \addplot[
        red,
        dashed,
        mark=none,
         thick
    ] coordinates {
(50,0.11056666666666665)
(100,0.0114)
(150,0.006900000000000001)
(200,0.0034)
    };

    \addplot[
        fill=blue!30,
        draw=none,
        fill opacity=0.3
    ] coordinates {
(50,0.16262674946709413)
(100,0.020840342392350233)
(150,0.0064930060189038255)
(200,0.00039999999999999996)
(200,-0.00019999999999999998)
(150,-0.003826339352237159)
(100,-0.005640342392350233)
(50,-0.05422674946709413)
    } \closedcycle;

    \addplot[
        fill=red!30,
        draw=none,
        fill opacity=0.3
    ] coordinates {
(50,0.2790078895177121)
(100,0.0271175061635108)
(150,0.018111155159036916)
(200,0.010101243665668834)
(200,-0.0033012436656688336)
(150,-0.0043111551590369135)
(100,-0.004317506163510801)
(50,-0.05787455618437881)
    } \closedcycle;

    \end{axis}
\end{tikzpicture}
  \vspace{-0.6cm}
   \caption{\textsc{BDB-C}}
\end{subfigure}
~\hfill
\begin{subfigure}{0.15\textwidth}
  \centering
    \begin{tikzpicture}
    \begin{axis}[
        axis x line  = bottom,
        axis y line  = left,
        width=4cm, height=4cm,
        xlabel={Budget},
        ylabel={Normalized performance},
        grid=minor,
          ymax=0.00022,
        ymin=-0.00001,
        legend columns=2, 
        legend style={draw=none,fill=none,at={(1.0,1)}, anchor=south east, legend cell align=left}
    ]

    \addplot[
        blue,
        mark=none,
         thick
    ] coordinates {
(50,0.0002)
(100,0.00013333333333333334)
(150,0.0)
(200,0.0)
    };

    \addplot[
        red,
        dashed,
        mark=none,
         thick
    ] coordinates {
(50,0.00023333333333333333)
(100,6.666666666666667e-05)
(150,0.0)
(200,0.0)
    };

    \addplot[
        fill=blue!30,
        draw=none,
        fill opacity=0.3
    ] coordinates {
(50,0.0006)
(100,0.0004732679675728523)
(150,0.0)
(200,0.0)
(200,0.0)
(150,0.0)
(100,-0.0002066013009061856)
(50,-0.00019999999999999996)
    } \closedcycle;

    \addplot[
        fill=red!30,
        draw=none,
        fill opacity=0.3
    ] coordinates {
(50,0.0007921018204746738)
(100,0.0003161104924515961)
(150,0.0)
(200,0.0)
(200,0.0)
(150,0.0)
(100,-0.00018277715911826277)
(50,-0.00032543515380800713)
    } \closedcycle;

    \end{axis}
\end{tikzpicture}
  \vspace{-0.6cm}
   \caption{\textsc{DeepArch}}
\end{subfigure}
~\hfill
\begin{subfigure}{0.15\textwidth}
  \centering
    \begin{tikzpicture}
    \begin{axis}[
        axis x line  = bottom,
        axis y line  = left,
        width=4cm, height=4cm,
        xlabel={Budget},
        ylabel={Normalized performance},
        grid=minor,
        legend columns=2, 
        scaled y ticks={base 10:2},
        ymax=0.21,
        ymin=0.02,
        legend style={draw=none,fill=none,at={(1.0,1)}, anchor=south east, legend cell align=left}
    ]

    \addplot[
        blue,
        mark=none,
         thick
    ] coordinates {
(50,0.20293333333333333)
(100,0.06396666666666666)
(150,0.041966666666666666)
(200,0.026833333333333334)
    };

    \addplot[
        red,
        dashed,
        mark=none,
         thick
    ] coordinates {
(50,0.14539999999999997)
(100,0.0869)
(150,0.0656)
(200,0.04466666666666667)
    };

    \addplot[
        fill=blue!30,
        draw=none,
        fill opacity=0.3
    ] coordinates {
(50,0.3755629818748152)
(100,0.15011788385030383)
(150,0.1094008238539227)
(200,0.07810131120993341)
(200,-0.024434644543266743)
(150,-0.025467490520589373)
(100,-0.022184550516970505)
(50,0.030303684791851443)
    } \closedcycle;

    \addplot[
        fill=red!30,
        draw=none,
        fill opacity=0.3
    ] coordinates {
(50,0.2303386445225807)
(100,0.17418701698038108)
(150,0.12872822929033678)
(200,0.10021024315296155)
(200,-0.010876909819628217)
(150,0.002471770709663243)
(100,-0.0003870169803810858)
(50,0.06046135547741924)
    } \closedcycle;

    \end{axis}
\end{tikzpicture}
        \vspace{-0.6cm}
   \caption{\textsc{PostgreSQL}}
\end{subfigure}


\begin{subfigure}{0.15\textwidth}
  \centering
    \begin{tikzpicture}
    \begin{axis}[
        axis x line  = bottom,
        axis y line  = left,
        width=4cm, height=4cm,
        xlabel={Budget},
        ylabel={Normalized performance},
        scaled y ticks={base 10:2},
        grid=minor,
        legend columns=2, 
        ymax=0.19,
        ymin=0.04,
        legend style={draw=none,fill=none,at={(1.0,1)}, anchor=south east, legend cell align=left}
    ]

    \addplot[
        blue,
        mark=none,
         thick
    ] coordinates {
(50,0.13796666666666668)
(100,0.0566)
(150,0.04926666666666667)
(200,0.0421)
    };

    \addplot[
        red,
        dashed,
        mark=none,
         thick
    ] coordinates {
(50,0.1822666666666667)
(100,0.08003333333333335)
(150,0.06843333333333333)
(200,0.05290000000000001)
    };

    \addplot[
        fill=blue!30,
        draw=none,
        fill opacity=0.3
    ] coordinates {
(50,0.26316929798441846)
(100,0.09369321591162819)
(150,0.07863550881083178)
(200,0.05836109877386314)
(200,0.02583890122613686)
(150,0.01989782452250155)
(100,0.01950678408837181)
(50,0.012764035348914876)
    } \closedcycle;

    \addplot[
        fill=red!30,
        draw=none,
        fill opacity=0.3
    ] coordinates {
(50,0.32751838687067053)
(100,0.14217053814834107)
(150,0.12671694071044143)
(200,0.07911494484703971)
(200,0.02668505515296031)
(150,0.010149725956225243)
(100,0.017896128518325602)
(50,0.037014946462662845)
    } \closedcycle;

    \end{axis}
\end{tikzpicture}
      \vspace{-0.6cm}
   \caption{\textsc{JavaGC}}
\end{subfigure}
~\hfill
\begin{subfigure}{0.15\textwidth}
  \centering
    \begin{tikzpicture}
    \begin{axis}[
        axis x line  = bottom,
        axis y line  = left,
        width=4cm, height=4cm,
        xlabel={Budget},
        ylabel={Normalized performance},
        grid=minor,
        legend columns=2, 
        ymax=0.0082,
        ymin=-0.0002,
        legend style={draw=none,fill=none,at={(1.0,1)}, anchor=south east, legend cell align=left}
    ]

    \addplot[
        blue,
        mark=none,
         thick
    ] coordinates {
(50,0.005466666666666667)
(100,0.0)
(150,0.0)
(200,0.0)
    };

    \addplot[
        red,
        dashed,
        mark=none,
         thick
    ] coordinates {
(50,0.008)
(100,0.0021333333333333334)
(150,0.0005333333333333334)
(200,0.0005333333333333334)
    };

    \addplot[
        fill=blue!30,
        draw=none,
        fill opacity=0.3
    ] coordinates {
(50,0.013228682441508207)
(100,0.0)
(150,0.0)
(200,0.0)
(200,0.0)
(150,0.0)
(100,0.0)
(50,-0.002295349108174872)
    } \closedcycle;

    \addplot[
        fill=red!30,
        draw=none,
        fill opacity=0.3
    ] coordinates {
(50,0.016610071621846897)
(100,0.0075722874811656364)
(150,0.003405421230471735)
(200,0.003405421230471735)
(200,-0.002338754563805068)
(150,-0.002338754563805068)
(100,-0.0033056208144989696)
(50,-0.0006100716218468989)
    } \closedcycle;

    \end{axis}
\end{tikzpicture}
      \vspace{-0.6cm}
   \caption{\textsc{Storm}}
\end{subfigure}
~\hfill
\begin{subfigure}{0.15\textwidth}
  \centering
    \begin{tikzpicture}
    \begin{axis}[
        axis x line  = bottom,
        axis y line  = left,
        width=4cm, height=4cm,
        xlabel={Budget},
        ylabel={Normalized performance},
        grid=minor,
        legend columns=2, 
        ymax=0.265,
        ymin=0.197,
        legend style={draw=none,fill=none,at={(1.0,1)}, anchor=south east, legend cell align=left}
    ]

    \addplot[
        blue,
        mark=none,
         thick
    ] coordinates {
(50,0.24603333333333338)
(100,0.21983333333333338)
(150,0.20376666666666673)
(200,0.19770000000000007)
    };

    \addplot[
        red,
        dashed,
        mark=none,
         thick
    ] coordinates {
(50,0.26026666666666665)
(100,0.22446666666666673)
(150,0.2146666666666667)
(200,0.20366666666666672)
    };

    \addplot[
        fill=blue!30,
        draw=none,
        fill opacity=0.3
    ] coordinates {
(50,0.3191070328123905)
(100,0.28264685806239864)
(150,0.264193307222707)
(200,0.2498594670218169)
(200,0.14554053297818323)
(150,0.1433400261106265)
(100,0.15701980860426815)
(50,0.17295963385427623)
    } \closedcycle;

    \addplot[
        fill=red!30,
        draw=none,
        fill opacity=0.3
    ] coordinates {
(50,0.3446400168562983)
(100,0.2878352884145178)
(150,0.2734302230222478)
(200,0.2595500770453789)
(200,0.14778325628795455)
(150,0.1559031103110856)
(100,0.16109804491881569)
(50,0.17589331647703496)
    } \closedcycle;

    \end{axis}
\end{tikzpicture}
      \vspace{-0.6cm}
   \caption{\textsc{x264}}
\end{subfigure}
~\hfill
\begin{subfigure}{0.15\textwidth}
  \centering
    \begin{tikzpicture}
    \begin{axis}[
        axis x line  = bottom,
        axis y line  = left,
        width=4cm, height=4cm,
        xlabel={Budget},
        ylabel={Normalized performance},
        grid=minor,
        legend columns=2, 
        ymax=0.47,
        ymin=0.23,
        legend style={draw=none,fill=none,at={(1.0,1)}, anchor=south east, legend cell align=left}
    ]

    \addplot[
        blue,
        mark=none,
         thick
    ] coordinates {
(50,0.39019999999999994)
(100,0.3071333333333334)
(150,0.26306666666666667)
(200,0.23616666666666664)
    };

    \addplot[
        red,
        dashed,
        mark=none,
         thick
    ] coordinates {
(50,0.4612666666666667)
(100,0.33190000000000003)
(150,0.2882)
(200,0.2500666666666667)
    };

    \addplot[
        fill=blue!30,
        draw=none,
        fill opacity=0.3
    ] coordinates {
(50,0.5422991781700348)
(100,0.4423466791572161)
(150,0.40522524791388587)
(200,0.3745398536968456)
(200,0.09779347963648768)
(150,0.12090808541944748)
(100,0.17191998750945076)
(50,0.23810082182996511)
    } \closedcycle;

    \addplot[
        fill=red!30,
        draw=none,
        fill opacity=0.3
    ] coordinates {
(50,0.6083688835847751)
(100,0.4636121482627932)
(150,0.4329916664268585)
(200,0.3847727500069854)
(200,0.11536058332634802)
(150,0.1434083335731415)
(100,0.2001878517372069)
(50,0.3141644497485583)
    } \closedcycle;

    \end{axis}
\end{tikzpicture}
      \vspace{-0.6cm}
   \caption{\textsc{Redis}}
\end{subfigure}
~\hfill
\begin{subfigure}{0.15\textwidth}
  \centering
    \begin{tikzpicture}
    \begin{axis}[
        axis x line  = bottom,
        axis y line  = left,
        width=4cm, height=4cm,
        xlabel={Budget},
        ylabel={Normalized performance},
        grid=minor,
        legend columns=2,
        ymax=0.022,
        ymin=0.003,
        legend style={draw=none,fill=none,at={(1.0,1)}, anchor=south east, legend cell align=left}
    ]

    \addplot[
        blue,
        mark=none,
         thick
    ] coordinates {
(50,0.021566666666666668)
(100,0.009333333333333334)
(150,0.003433333333333333)
(200,0.003433333333333333)
    };

    \addplot[
        red,
        dashed,
        mark=none,
         thick
    ] coordinates {
(50,0.01856666666666667)
(100,0.009766666666666668)
(150,0.007333333333333334)
(200,0.005866666666666667)
    };

    \addplot[
        fill=blue!30,
        draw=none,
        fill opacity=0.3
    ] coordinates {
(50,0.04966374045657503)
(100,0.024251629333733496)
(150,0.011605168432556084)
(200,0.011605168432556084)
(200,-0.004738501765889417)
(150,-0.004738501765889417)
(100,-0.005584962667066828)
(50,-0.006530407123241697)
    } \closedcycle;

    \addplot[
        fill=red!30,
        draw=none,
        fill opacity=0.3
    ] coordinates {
(50,0.04037749476241356)
(100,0.024044115863058702)
(150,0.020819952216095942)
(200,0.018739722075370367)
(200,-0.007006388742037032)
(150,-0.006153285549429273)
(100,-0.004510782529725366)
(50,-0.0032441614290802286)
    } \closedcycle;

    \end{axis}
\end{tikzpicture}
      \vspace{-0.6cm}
   \caption{\textsc{HSQLDB}}
\end{subfigure}
~\hfill
\begin{subfigure}{0.15\textwidth}
  \centering
    \begin{tikzpicture}
    \begin{axis}[
        axis x line  = bottom,
        axis y line  = left,
        width=4cm, height=4cm,
        xlabel={Budget},
        ylabel={Normalized performance},
        grid=minor,
        ymax=0.017,
        ymin=-0.001,
        legend columns=2, 
        legend style={draw=none,fill=none,at={(1.0,1)}, anchor=south east, legend cell align=left}
    ]

    \addplot[
        blue,
        mark=none,
         thick
    ] coordinates {
(50,0.011466666666666668)
(100,0.0)
(150,0.0)
(200,0.0)
    };

    \addplot[
        red,
        dashed,
        mark=none,
         thick
    ] coordinates {
(50,0.016733333333333333)
(100,0.0035)
(150,0.0)
(200,0.0)
    };

    \addplot[
        fill=blue!30,
        draw=none,
        fill opacity=0.3
    ] coordinates {
(50,0.02928001996006866)
(100,0.0)
(150,0.0)
(200,0.0)
(200,0.0)
(150,0.0)
(100,0.0)
(50,-0.006346686626735322)
    } \closedcycle;

    \addplot[
        fill=red!30,
        draw=none,
        fill opacity=0.3
    ] coordinates {
(50,0.040092747329027276)
(100,0.009989735485107747)
(150,0.0)
(200,0.0)
(200,0.0)
(150,0.0)
(100,-0.0029897354851077473)
(50,-0.00662608066236061)
    } \closedcycle;

    \end{axis}
\end{tikzpicture}
      \vspace{-0.6cm}
   \caption{\textsc{LLVM}}
\end{subfigure}

\vspace{-0.2cm}
\caption{Ablating causally purified rules over 30 runs (smaller performance is better). \textcolor{blue}{\textbf{---}} and \textcolor{red}{\textbf{- - -}} denote \tuner~and \texttt{w/o Rules}, respectively. Raw data can be accessed at \textcolor{blue}{\texttt{\href{https://github.com/ideas-labo/PromiseTune/blob/main/RQs/RQ2/rq2.pdf}{https://github.com/ideas-labo/PromiseTune/blob/main/RQs/RQ2/rq2.pdf}}}.}
     \label{fig:rq2}
     \vspace{-0.3cm}
\end{figure*}

\subsubsection{Method}

We conduct ablation analysis in \textbf{RQ2}. The key designs of \tuner~that impact the performance are the interrelated rule generation and purification via causal inference, which cannot be separated. Therefore, we assess \tuner~against its variant that the rules generator and rules causality purifier are turned off\footnote{This is essentially the same as only turning off the causality purifier, as using all the learned rules simply means that we sample in the entire configuration landscape.} (denoted as \texttt{w/o Rules}), i.e., the tuning is not guided by any rule but a Random Forest-based Bayesian optimization. 




\subsubsection{Results}

The results from Figure~\ref{fig:rq2} clearly indicate the necessity of rules and their causal purification: \tuner~leads to generally better and more stable performance; in 7 out of 12 systems, it has better results on all budgets while on the others (e.g., \textsc{PostgreSQL}), it has inferior result on at most one budget only. Those demonstrate that, regardless of the systems or budgets, the rules, after purification, can effectively guide the tuning towards the promising regions, hence finding better configurations that would otherwise be difficult to find. Therefore, we conclude:


\begin{quotebox}
   \noindent
   \textit{\textbf{RQ2:} The causally purified rules play an important role in \tuner~that significantly contributes to its success, achieving better and more stable performance.}
\end{quotebox}



\subsection{RQ3: Sensitivity to $l$}
\label{sec:rq3}


\subsubsection{Method}

The most crucial parameter in \tuner~is $l$ for the Random Forest that learns the rules. $l$ sets the minimal number of leaves, hence affecting the smallest number of rules learned, impacting both the performance and explainability of \tuner. The bigger $l$ can encourage more leaves, hence the sub-trees tend to be flat, leading to simpler/shorter rules. In contrast, smaller $l$ makes deeper sub-trees, leading to complex/longer rules. To study the sensitivity of \tuner~to $l$ in \textbf{RQ3}, we test \tuner~under different $l$ values: $\{5,10,15,20\}$. We report on the normalized mean and standard deviation of performance, together with the number of explainable rules, for all systems over 30 runs under each of the three budget settings.

\subsubsection{Results}

\begin{figure}[t!]
\centering
\subfloat[Performance]{
\begin{tikzpicture}
    \begin{axis}[
        axis x line = bottom,
        axis y line = left,
        width=5cm, height=5cm, 
        xlabel={$l$ value},
        ylabel={Normalized performance},
        ymin=0.045, ymax=0.133, 
        grid=minor,
        legend columns=1,
        y label style={at={(-0.05,0.5)}},
        legend style={draw=none,fill=none,at={(1.1,0.6)}, anchor=south east, legend cell align=left}
    ]

    \addplot[
        orange,
        mark=none,
        ultra thick
    ] coordinates {(5,0.1139) (10,0.1112) (15,0.1258) (20,0.1289)};
    \addlegendentry{$B=50$}
    
    \addplot[
        blue,
        mark=none,
        ultra thick
    ] coordinates {(5,0.0689) (10,0.0647) (15,0.0647) (20,0.0678)};
    \addlegendentry{$B=100$}

    \addplot[
        red,
        mark=none,
        ultra thick
    ] coordinates {(5,0.0582) (10,0.0538) (15,0.0534) (20,0.0599)};
    \addlegendentry{$B=150$}

    \addplot[
        teal,
        mark=none,
        ultra thick
    ] coordinates {(5,0.0536) (10,0.0487) (15,0.0481) (20,0.0529)};
    \addlegendentry{$B=200$}

    \addplot[
        fill=yellow!30,
        draw=none,
        fill opacity=0.3
    ] coordinates {(5,0.022) (10,0.0274) (15,0.0262) (20,0.037) (20,0.2207) (15,0.2255) (10,0.1951) (5,0.2058)} \closedcycle;
    
    \addplot[
        fill=blue!30,
        draw=none,
        fill opacity=0.3
    ] coordinates {(5,0.0249) (10,0.024) (15,0.0091) (20,0.0142) (20,0.1214) (15,0.1203) (10,0.1054) (5,0.1129)} \closedcycle;

    \addplot[
        fill=red!30,
        draw=none,
        fill opacity=0.3
    ] coordinates {(5,0.0217) (10,0.0181) (15,0.0149) (20,0.0104) (20,0.1095) (15,0.092) (10,0.0895) (5,0.0948)} \closedcycle;

    \addplot[
        fill=teal!30,
        draw=none,
        fill opacity=0.3
    ] coordinates {(5,0.0175) (10,0.0169) (15,0.0148) (20,0.0096) (20,0.0961) (15,0.0814) (10,0.0805) (5,0.0897)} \closedcycle;

    \end{axis}
\end{tikzpicture}
}
\subfloat[Rule count]{
\begin{tikzpicture}
    \begin{axis}[
      axis x line  = bottom,
            axis y line  = left  ,
        width=5cm, height=5cm, 
        xlabel={$l$ value},
        ylabel={$\#$ of explainable rules},
         y label style={at={(-0.05,0.5)}},
        grid=minor,
        legend style={draw=none,fill=none,at={(1.1,0.6)}, anchor=south east, legend cell align=left}
    ]

    \addplot[
        orange,
        mark=none,
        ultra thick
    ] coordinates {
(5,7.7)
(10,4.863333333333333)
(15,2.726666666666667)
(20,0.0)

    };
    \addlegendentry{$B=50$}

    \addplot[
        blue,
        mark=none,
        ultra thick
    ] coordinates {
(5,13.166666666666666)
(10,10.956666666666667)
(15,5.323333333333333)
(20,2.9233333333333333)

    };
    \addlegendentry{$B=100$}

    \addplot[
        red,
        mark=none,
        ultra thick
    ] coordinates {
(5,17.473333333333336)
(10,17.163333333333334)
(15,6.92)
(20,5.966666666666667)
    };
     \addlegendentry{$B=150$}

    \addplot[
        teal,
        mark=none,
        ultra thick
    ] coordinates {
(5,22.163333333333334)
(10,20.813333333333336)
(15,10.543333333333333)
(20,6.5200000000000005)

    };
    \addlegendentry{$B=200$}

    \addplot[
        fill=orange!30, 
        draw=none,
        fill opacity=0.3 
    ] coordinates {
(5,5.582779062483466)
(10,2.7695694410413276)
(15,1.8510726304276144)
(20,0.0)
(20,0.0)
(15,3.6022607029057196)
(10,6.957097225625339)
(5,9.817220937516534)

    } \closedcycle;

    \addplot[
        fill=blue!30, 
        draw=none,
        fill opacity=0.3 
    ] coordinates {
(5,10.155312289074692)
(10,7.8636108211213)
(15,3.4542921510182536)
(20,1.7838906658727403)
(20,4.062776000793926)
(15,7.192374515648413)
(10,14.049722512212034)
(5,16.178021044258642)

    } \closedcycle;

    \addplot[
        fill=red!30, 
        draw=none,
        fill opacity=0.3 
    ] coordinates {
(5,14.10341539512644)
(10,13.150869383628322)
(15,4.721688274407629)
(20,4.03046495769186)
(20,7.902868375641473)
(15,9.11831172559237)
(10,21.175797283038346)
(5,20.843251271540232)
    } \closedcycle;

    \addplot[
        fill=teal!30, 
        draw=none,
        fill opacity=0.3 
    ] coordinates {
(5,18.029990166530098)
(10,16.053403208058512)
(15,7.948931654670568)
(20,4.341221040360896)
(20,8.698778959639105)
(15,13.137735011996098)
(10,25.57326345860816)
(5,26.29667650013657)

    } \closedcycle;

    \end{axis}
\end{tikzpicture}
}

\caption{Sensitivity of \tuner~to parameter $l$ over all systems. The smaller the normalized performance, the better.}
\label{fig:rq3}
\end{figure}
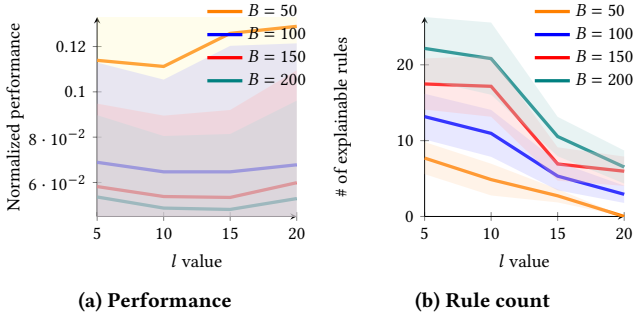

As in Figure~\ref{fig:rq3}a, we can observe that $l \in [10,15]$ leads to the generally optimal outcomes across the budgets ($l=10$ is the best when $B=50$)---neither too high nor too low $l$ is ideal. It is easy to understand that a bigger $l$ can result in too simple/short rules; hence, after purification, the remaining rules are hardly useful, providing limited guidance. In contrast, it might seem counter-intuitive to see that a smaller $l$ also leads to performance degradation. The reason is that decreasing the $l$ to a too-low value can be risky in creating too complex/long rules, which might incorrectly constrain the explored regions at the tuning, especially at the earlier stage where the data used to perform causal inference is limited. 

For the number of rules generated by \tuner~for explainability, in Figure~\ref{fig:rq3}b, there is a clear trade-off: a bigger $l$ can cause many simple/short rules to be eliminated at purification, which is easier for the comprehension of explainability but might not be informative. On the other hand, a smaller $l$ will preserve many complex/long rules, but can easily create cognitive fragility to the explainability. As such, we set $l=10$ as the default. Overall, we say

\begin{quotebox}
   \noindent
   \textit{\textbf{RQ3:} \tuner~is sensitives to $l$ for which $l=10$ tends to be the safe setting, achieving the generally acceptable performance while balancing the comprehensiveness and cognitive overhead in explainability.}
\end{quotebox}



\subsection{RQ4: Explainability Case Study}
\label{sec:rq4}



\subsubsection{Method}

To assess the explainability of \tuner~in \textbf{RQ4}, we conduct a case study (at $B=200$) on a randomly chosen system and compare it with \texttt{Unicorn}~\cite{DBLP:conf/eurosys/IqbalKJRJ22}, another explainable tool for configurable systems using causal inference at the option level. We firstly check the number of explainable rules with different $k$ values ($k \in \{5,10,...,50\}$). We then examine the explainable rules returned when $k=10$: all causally purified rules that cover the top 10\% performing configurations found in the tuning. 


\subsubsection{Results}

We apply both \tuner~and \texttt{Unicorn} on \textsc{x264}, which has 17 options to tune. From Table~\ref{tab:explain}, as expected, bigger $k$ leads to gradually more explainable rules. Notably, when $k=10$, we see that both tuners provide the following information:

\begin{itemize}
    \item \texttt{Unicorn} lists the three most influential options for performance out of the 17 options.
    \item The explainable rules contain six options, as contained in the rules, out of the 17 options.
\end{itemize}

Following Section~\ref{sec:exp-thoery}, when explaining at the option level, both tuners recommend \texttt{Crf}, \texttt{Seek}, and \texttt{Ipratio} as the keys, while \tuner~additionally includes three others\footnote{We have verified that the additional three can indeed influence the performance, hence they are the false negatives for \texttt{Unicorn}.}. Importantly, we see that \texttt{Crf}, \texttt{Seek}, and \texttt{Ipratio} are more commonly involved in the rules, suggesting their higher importance. Further, \tuner~confirms that those options are causally important for finding promising configurations while \texttt{Unicorn} only suggests that those options can causally impact the performance.

\begin{table}[t]

\caption{Sensitivity to $k$ together with the explainable outcomes returned by \tuner~($k=10$) and \texttt{Unicorn} for \textsc{x264}.}
\label{tab:explain}
\begin{adjustbox}{width=\linewidth, center}
  \centering
  \begin{tabular}{l||p{5.5cm}|p{3cm}}
    \toprule
     \textbf{$k\rightarrow\#$ rules}&
    \textbf{\tuner~(explainable rules at $k=10$)} & \textbf{\texttt{Unicorn} (explainable options)}\\  \midrule

 $k=5\rightarrow$10 & ${R}_{1}=\langle$\texttt{Crf>33}, \texttt{Seek<541}$\rangle$& \texttt{Ipratio}\\
  $k=10\rightarrow$10 &${R}_{2}=\langle$\texttt{Crf>36}, \texttt{Seek<541}$\rangle$& \texttt{Crf}\\
  $k=15\rightarrow$11 &${R}_{3}=\langle$\texttt{Crf>26}, \texttt{Seek<523}$\rangle$& \texttt{Seek}\\
  $k=20\rightarrow$14 &${R}_{4}=\langle$\texttt{Crf>36}, \texttt{Ipratio<0}$\rangle$& \\
  $k=25\rightarrow$14 &${R}_{5}=\langle$\texttt{Crf>26}, \texttt{Qp>30}$\rangle$& \\
  $k=30\rightarrow$14 &${R}_{6}=\langle$\texttt{Crf>26}, \texttt{B\_bias>15}, \texttt{Scenecut>44}$\rangle$& \\
  $k=35\rightarrow$14 &${R}_{7}=\langle$\texttt{Crf>36}, \texttt{Ipratio>0}$\rangle$& \\
  $k=40\rightarrow$14 &${R}_{8}=\langle$\texttt{Crf>26}, \texttt{Qp<30}$\rangle$& \\
  $k=45\rightarrow$14 &${R}_{9}=\langle$\texttt{Crf<36}, \texttt{Seek<627}, \texttt{Qp>20}$\rangle$& \\
  $k=50\rightarrow$14 &${R}_{10}=\langle$\texttt{Crf<36}, \texttt{Seek<731}, \texttt{B\_bias>-16}$\rangle$& \\

    \bottomrule
  \end{tabular}
\end{adjustbox}
\end{table}

Importantly, \tuner~can further explain the following at the landscape level that has not been covered in \texttt{Unicorn}:

\begin{itemize}
    \item The interaction between \texttt{Crf} and \texttt{Seek} are the most likely leading to the promising regions in the configuration landscape. Therefore, a tuner or future configurable system design should take their co-adjustment into account.
    \item The most common overlapping(s) that is covered by the most rules can highlight the most important promising regions in the landscape, providing spatial insights for future analysis. For example, the most common overlapping(s) is: \texttt{Crf} $>36$ is covered by $R_1$--$R_8$; \texttt{Seek} $<523$ is covered by $R_1$--$R_3$, $R_9$, and $R_{10}$; \texttt{Qp} $>30$ is covered by $R_5$ and $R_9$; $20<$ \texttt{Qp} $<30$ is covered by $R_8$ and $R_9$; together with \texttt{B\_bias} $>15$ and \texttt{Scenecut} $>44$ (\texttt{Ipratio} has no most common overlap among the rules), suggesting that two specific, most promising regions are: $\langle$\texttt{Crf} $>36$, \texttt{Seek} $<523$, \texttt{Qp} $>30$, \texttt{B\_bias} $>15$, \texttt{Scenecut} $>44\rangle$ and $\langle$\texttt{Crf} $>36$, \texttt{Seek} $<523$, $20<$ \texttt{Qp} $<30$, \texttt{B\_bias} $>15$, \texttt{Scenecut} $>44\rangle$ (the absent options are unbounded).
    
\end{itemize}

We verified that the two most promising regions can indeed bound most of the good configurations. Thus, we conclude that:

\begin{quotebox}
   \noindent
   \textit{\textbf{RQ4:} Compared with option level explainable tuner \texttt{Unicorn}, \tuner~is not only able to explain option importance, but can also provide additional explanation of spatial information at the landscape level, i.e., option interactions and the most promising regions.}
\end{quotebox}




\section{Discussion}
\label{sec:discussion}

\subsection{Why dose \tuner~Work?}

\begin{figure}[t!]
\centering
\subfloat[\tuner~(darker red is better)]{
\includegraphics[width=.495\linewidth]{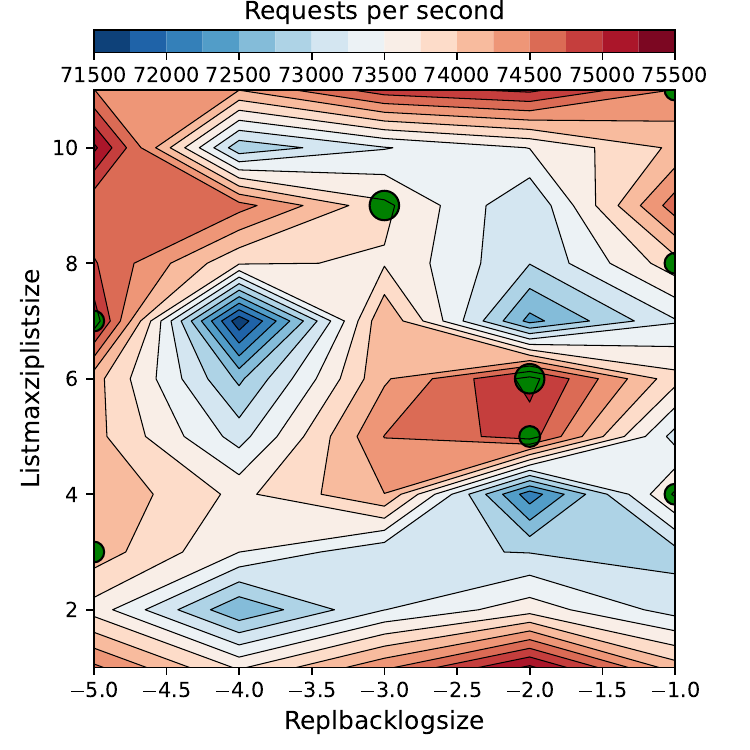}
\label{fig: Redis_ours}
}
\subfloat[\texttt{MBO}~(darker red is better)]{
\includegraphics[width=.495\linewidth]{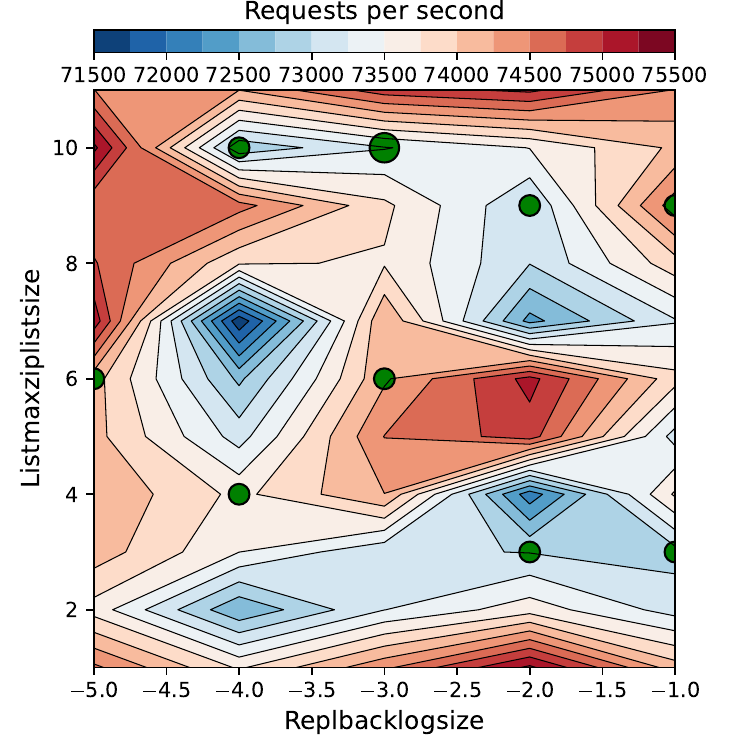}
\label{fig: Redis_MBO}
}

\caption{The explored configurations within the last $10$\% budget by \tuner~and \texttt{MBO} (the second best tuner in this case) in one run when tuning \textsc{Redis}. \quart{42.03}{57.97}{71.01}{100} denotes the position of the configuration visited by the tuner in the landscape. A bigger \quart{42.03}{57.97}{71.01}{100} means that the configuration with the same values of the options has been visited more often.}
\label{fig:discussion}
\end{figure}

The key motivation behind \tuner~is that it aims to approximate the promising regions, and only explore within those regions in the tuning, mitigating the issues caused by the trade-off between jumping out of local optima and fully utilizing the budgets for better configurations. To understand how this is achieved, Figure~\ref{fig:discussion} shows the example landscapes of a system. We see that \tuner~has successfully found configurations close to the promising regions; the \texttt{MBO}, in contrast, finds points that are more spread apart. This explains why \tuner~outperforms the others in general---the approximated promising regions, represented by rules, are effective in guiding the tuning to concentrate on those regions, and hence considerably improve the budget utilization.

\subsection{What Implications can the Explainable Rules from \tuner~Bring?}
\label{sec:exp-why}

The key explainability that \tuner~offers is the rich information on the spatial aspect of the configuration landscape, since the finally produced/extracted rules from \tuner~bound the likely promising regions. Such information provides several additional insights and complements the other explainable approaches that focus on options. The implications include the following.

\textbf{For Researchers on System-specific Tuner Design:} it is not hard to expect that those promising regions reflected by the explainable rules, once identified and verified by \tuner, can then be used to \textit{specialize a tuner particularly for the system under tuning}. For example, in the subsequent tuning, 

    \begin{itemize}
        \item the search operator can be designed to target around the most important region reflected by the most common overlapping(s) of the explainable rules from \tuner, hence using the budget more precisely;
        \item the options to be considered can be reduced to those only present in the rules. While the existing explainable tuners can also achieve similar results, \tuner~produce something different: it only leaves those options, which are likely to be helpful in finding the promising regions and configurations, in the rules; whereas existing explainable tuners are mainly concerned about the most influential options, e.g., an option is said important even though it might only explore the low-performing regions~\cite{DBLP:conf/eurosys/IqbalKJRJ22,DBLP:journals/corr/abs-2402-05399}.
    \end{itemize}

    While the specialization can make a tuner less general, in a practical scenario, having such a specialized tuner targeting the concerned system can often lead to considerably better outcomes.

\textbf{For Developers on Configurable System Design:} Those explainable rules and their promising regions can also help to analyze the behaviors of the systems in a fine-grained manner: they do not only show the important options for finding good configurations but can also reflect on the promising configuration regions bounded by particular values of the options. This can help developers \textit{better engineer configurable systems that are ``easier'' to tune} from various aspects: 

    \begin{itemize}
        \item to merge some options in future releases of the system that often interact together (as indicated by different rules) to form promising configurations;
        \item to provide more comprehensive manuals/documentation on how to set the values of the options;
        \item to refactor the system code, adding constraints to the values of certain options, and hence only the values within the bounds of the rules/promising regions can be set.
    \end{itemize}
    

All of the above can only be achieved by explaining in a fine-grained manner at the landscape level in \tuner~as opposed to the coarse-grained explainability at the options level.

\section{Threats to Validity}
\label{sec:threats}

\textbf{Internal threats} to validity are related to the parameters used. For \tuner, we set the parameter $l=10$---an appropriate choice verified in \textbf{RQ3}, serving as a ``rule-of-thumb'' that yields generally favourable outcomes. As for $k$, it entirely depends on how many explainable rules one wishes to examine. For the settings of the others, we follow the default and what has been used in prior work~\cite{DBLP:conf/lion/HutterHL11,DBLP:conf/sigsoft/ShahbazianKBM20,DBLP:conf/kdd/LiSZCJLJG0Y0021,DBLP:conf/sc/BehzadLHBPAKS13,DBLP:conf/sigsoft/ShahbazianKBM20,DBLP:journals/tse/Nair0MSA20,DBLP:conf/eurosys/IqbalKJRJ22,DBLP:conf/icse/0003XC021,DBLP:conf/kbse/ZhuH23,DBLP:conf/sigmod/AkenPGZ17,DBLP:journals/pvldb/KanellisDKMCV22}. We use diverse budgets tailored to fit our needs, considering the most commonly used values and the computational resources we have. Yet, the optimal settings, especially for $l$, might need to be adjusted for each system. 


\textbf{Construct threats} to validity may be incurred by the metrics used. In this work, quantitatively, we use the performance optimized by different tuners over the systems, which is the most intuitive and concerned metric. For the specific explainability provided by \tuner, we provide a qualitative case study to evaluate the rich information in the explainable rules. However, unintentional small errors, such as minor programming issues, might be possible.

\textbf{External threats} could be raised from the subject systems and data samples used. To mitigate this, we cover 12 systems of different characteristics and four budget sizes, leading to 48 cases, in each of which \tuner~is evaluated against 11 state-of-the-art tuners from different research communities. Indeed, we acknowledge that comparing more systems and tuners might prove more fruitful.




\section{Related Work}
\label{sec:related_work}

\subsection{Configuration Performance Learning}

There has been much work on learning the correlation between configuration options and performance~\cite{DBLP:journals/tse/ChenB17,gong2024dividable,DBLP:journals/pacmse/Gong024,8329264,9401991}. For example, Gong and Chen~\cite{DBLP:conf/sigsoft/Gong023,gong2024dividable} propose \texttt{DaL}, which leverages multiple neural networks and sample divisions to create local models for predicting configurations, together with the online extension~\cite{DBLP:conf/icse/XiangChen26}. Other works have built models that exploit data collected from different environments, e.g., \texttt{SeMPL}~\cite{DBLP:journals/pacmse/Gong024} and \texttt{BEETLE}~\cite{8329264}. White box approaches also exist. For example, \texttt{Comprex}~\cite{9401991} builds local models by analyzing the configuration code, based on which the structural information of the code can be explained. 

Yet, the above emphasizes modeling, i.e., predicting performance for a given configuration while \tuner~targets optimization, i.e., finding the best configuration via tuning. As such, those models are complementary to \tuner. Further, unlike those, the causal model learned in \tuner~focuses on the relationships between featurized configuration rules and performance.



\subsection{Tuning with or without Models}




Configuration tuning has been tackled using model-free heuristics, i.e., the search is guided solely on system measurements~\cite{DBLP:journals/tsc/ChenB17,DBLP:journals/tosem/ChenLBY18,DBLP:conf/wcre/Chen22,DBLP:conf/sigmetrics/SullivanSP04, DBLP:conf/cloud/ZhuLGBMLSY17, DBLP:journals/tse/ChenNKM19, DBLP:conf/sigsoft/0001L24,DBLP:journals/corr/abs-2112-07303,DBLP:conf/sigsoft/0001L21,DBLP:conf/sc/BehzadLHBPAKS13,DBLP:conf/sigsoft/ShahbazianKBM20}. For example, \texttt{GA} has been widely used as the foundation in different tuners that leverage population of configurations to evolve for better ones~\cite{DBLP:conf/sc/BehzadLHBPAKS13,DBLP:conf/sigsoft/ShahbazianKBM20}. \texttt{MMO}~\cite{DBLP:conf/sigsoft/0001L24,DBLP:journals/corr/abs-2112-07303,DBLP:conf/sigsoft/0001L21} is a new multi-objectivization optimization model to tune a single performance objective by adopting the multi-objective version of the \texttt{GA}, although it assumes the presence of multiple performance metrics.

In contrast, model-based tuners use a surrogate performance model, paired with real measurements and other heuristics, to expedite the tuning~\cite{DBLP:conf/lion/HutterHL11, DBLP:journals/tse/Nair0MSA20, DBLP:conf/icse/0003XC021, DBLP:conf/sigmod/AkenPGZ17, DBLP:conf/kbse/ZhuH23}. Among others, \texttt{OtterTune}~\cite{DBLP:conf/sigmod/AkenPGZ17} uses the Gaussian Process as their surrogate model while \texttt{FLASH}~\cite{DBLP:journals/tse/Nair0MSA20} uses a decision tree as the surrogate model to accelerate the search. Some other tuners consolidate the operators during the tuning, e.g., \texttt{BOCA}~\cite{DBLP:conf/icse/0003XC021} leverages Random Forest to identify the most important configuration options to serve as the key in the tuning and equip its sampling with a decay function, gradually reducing the use of those non-important options. Others use reinforcement learning~\cite{DBLP:conf/sigmod/ZhangLZLXCXWCLR19, DBLP:conf/sigmod/CaiLZZZLLCYX22} and Large Language Model (LLM) to assist the tuning~\cite{DBLP:journals/corr/abs-2408-02213,DBLP:journals/pvldb/LaoWLWZCCTW24}.

However, the above tuners all have no knowledge about the potentially promising regions, and hence they rely mainly on ``trial-and-error'' to balance using the budget for jumping out from local optima (exploration) and for finding better ones based on explored good configurations (exploitation). Unlike existing tuners, \tuner~is designed to guide the tuning for searching within likely promising regions, hence relieving the above issue.



\subsection{Explainability in Configuration Tuning}

Recently, there have been a few studies~\cite{DBLP:journals/corr/abs-2402-05399,DBLP:conf/cloud/IqbalZARJ23,DBLP:conf/eurosys/IqbalKJRJ22} leverage causality in explaining configurations. Among others, \texttt{Cure}~\cite{DBLP:journals/corr/abs-2402-05399} filter out the causally irrelevant options to explain the configuration analysis; \texttt{CAMEO} \cite{DBLP:conf/cloud/IqbalZARJ23} conducts transfer learning through causal inference to explain the relationships across hardware environments. Yet, their purposes are to understand configuration performance learning while \tuner~seek to explain the system behaviors with spatial information from the landscape.

\texttt{Unicorn}~\cite{DBLP:conf/eurosys/IqbalKJRJ22} adopts causal inference to estimate the important options for analyzing, debugging, and tuning configuration, but it differs from \tuner~such that:

\begin{itemize}
    \item \texttt{Unicorn} uses causal inference at the option level while \tuner~adopts it at the landscape level via analyzing the rules, which reflect regions in the landscape.
    \item \texttt{Unicorn} provides explainability on the most important options. \tuner, in contrast, provides explainability with more spatial information, e.g., option interaction for promising configurations and the most promising region by extracting the most common overlap of explainable rules.
    \item \tuner~directly leverages the causally purified rules to guide the tuning in an iterative manner while \texttt{Unicorn} only use the most causally related options to alter configurations at the last iteration of tuning.
\end{itemize}


\section{Conclusion}
\label{sec:conclusion}

This paper presents \tuner---a tuner that guides the model-based tuning via the likely promising regions reflected by learned and causally purified rules, in which both the rules and performance model are dually updated on-the-fly. The approximated promising regions not only mitigate the difficult trade-off between exploration and exploitation but also provide rich spatial information to support the explainability of the hidden system characteristics. By comparing \tuner~with 11 state-of-the-art tuners under 12 systems and varying budgets, we show that \tuner~performs considerably better and more stable than the others, being ranked the best in 63\% of the cases while offering richer spatial explainability at the landscape level.

We envisage that the insights from this work can stimulate fruitful future research on configuration tuning, paving the way towards more domain knowledge-guided and explainable tuner designs.

\section*{Acknowledgment}
This work was supported by a NSFC Grant (62372084) and a UKRI Grant (10054084).

\balance
\bibliographystyle{ACM-Reference-Format}
\bibliography{references}

\end{document}